\documentclass[twocolumn,conference,10pt,a4paper]{IEEEtran}

\IEEEoverridecommandlockouts

\usepackage{cite}
\usepackage{amsmath,amssymb,amsfonts}
\usepackage{algorithmic}
\usepackage{graphicx}
\usepackage{textcomp}
\usepackage{afterpage}
\usepackage{xr}
\def\BibTeX{{\rm B\kern-.05em{\sc i\kern-.025em b}\kern-.08em
		T\kern-.1667em\lower.7ex\hbox{E}\kern-.125emX}}
\usepackage{tikz}
\usetikzlibrary{backgrounds}
\usetikzlibrary{spy}
\usetikzlibrary{calc,arrows.meta}
\usepackage{pgfplots}
\usetikzlibrary{plotmarks}
\usetikzlibrary{arrows.meta}
\usetikzlibrary{positioning,fit}
\usetikzlibrary{intersections}
\usetikzlibrary{patterns}
\usetikzlibrary{decorations.shapes}
\usepackage{float}

\usepgfplotslibrary{patchplots}
\usepgfplotslibrary{colormaps}

\usepackage{cancel}

\usepackage{pdfpages} 

\usepackage{multirow}
\usepackage{makecell}

\usepackage{algorithmic}
\usepackage[ruled,vlined]{algorithm2e}
\usepackage{setspace}
\SetKwRepeat{Do}{do}{while}
\SetKwRepeat{Until}{do}{until}

\pgfdeclarelayer{back1}
\pgfdeclarelayer{lay1}
\pgfdeclarelayer{back2}
\pgfdeclarelayer{lay2}
\pgfsetlayers{back2,lay2,back1,lay1,main}

\newcommand{\exportFigures}{true}
\newcommand{\exportFiguresAsPNG}{true}

\ifthenelse{\equal{\exportFigures}{true}}
{
	\usepgfplotslibrary{external}
	\tikzexternalize[prefix=compiled_tikz_figures/,optimize command away=\includepdf]
	\ifthenelse{\equal{\exportFiguresAsPNG}{true}}
	{
		\tikzset
		{   png export/.style={
				external/system call={
					pdflatex \tikzexternalcheckshellescape -halt-on-error --extra-mem-top=10000000 -interaction=batchmode -jobname "\image" "\texsource" && pdftops -eps "\image.pdf" && convert -density 700 -transparent white "\image.pdf" "\image.png"
		}}}
		\tikzset{png export}
	}
	{}
}
{}

\usepackage{url}

\usepackage{bm}
\usepackage{subcaption}
\usepackage[font=footnotesize]{caption}
\usepackage[normalem]{ulem}

\usepackage{mathtools, cuted}

\usepackage{acronym}

\usepackage{booktabs}

\usepackage[latin1]{inputenc}
\usepackage{tikz}
\usetikzlibrary{shapes,arrows}
\usetikzlibrary{arrows.meta}
\usetikzlibrary{positioning}

\usepackage{ellipsis}

\usetikzlibrary{calc}
\usetikzlibrary{decorations.pathreplacing,decorations.markings,shapes.geometric}
\usetikzlibrary{decorations.pathmorphing}
\usetikzlibrary{fit}
\usetikzlibrary{pgfplots.groupplots}

\usetikzlibrary{calc,arrows.meta}

\usetikzlibrary{backgrounds} 

\definecolor{green(pigment)}{rgb}{0.0, 0.65, 0.31}
\definecolor{frenchblue}{rgb}{0.0, 0.45, 0.73} 
\definecolor{mediumcandyapplered}{rgb}{0.89, 0.02, 0.17}

\usepackage[most]{tcolorbox}
\tcbuselibrary{breakable}
\tcbset{every box/.style={enhanced,breakable}}
\tcbset{colframe=black,colback=red!10,enhanced,breakable,sharp corners}

\usepackage{enumitem} 

\usepackage{notation}
\usepackage{adjustbox}

\definecolor{alex}{RGB}{51,183,150}
\definecolor{erik}{RGB}{235,134,52}

\newcommand{\ticked}{$\text{\rlap{$\checkmark$}}\square$}
\newcommand{\unticked}{{$\square$}}
\newcommand{\tick}[1]{\ifthenelse{#1=1}{\ticked}{\unticked}}

\newcommand{\rmv}{\hspace*{-.3mm}}

\renewcommand{\exp}[1]{\ensuremath{{e}^{#1}}}

\newcommand{\norm}[2]{\ensuremath{\lVert #1 \rVert^{#2}}}

\newcommand{\transp}{\ensuremath{^\mathsf{T}}}

\newcommand{\minus}{\rmv - \rmv}

\newcommand{\s}{\hspace*{0.5pt}}

\newcommand{\pd}{p_{\text{d}}}

\providecommand{\norm}[1]{\lVert#1\rVert}

\newcommand{\ist}{\hspace*{.3mm}}
\newcommand{\iist}{\hspace*{1mm}}

\newcommand{\nn}{\nonumber}

\DeclareMathOperator{\atantwo}{atan2}

\newcommand{\zd}{\ensuremath{{z_\mathrm{d}^{(j)}}_{\rmv\rmv\rmv\rmv\rmv\rmv\rmv  m,n}}}
\newcommand{\zu}{\ensuremath{{z_\mathrm{u}^{(j)}}_{\rmv\rmv\rmv\rmv\rmv\rmv\rmv m,n}}}
\newcommand{\zaoa}{\ensuremath{{z_\mathrm{\theta}^{(j)}}_{\rmv\rmv\rmv\rmv\rmv\rmv\rmv  m,n}}}
\newcommand{\zaod}{\ensuremath{{z_\mathrm{\vartheta}^{(j)}}_{\rmv\rmv\rmv\rmv\rmv\rmv\rmv  m,n}}}

\newcommand{\fa}{f_{\mathrm{fa}}(\V{z}_{m,n}^{(j)})}
\DeclareMathOperator{\bdiag}{blkdiag}

\newlength{\figureheight}
\newlength{\figurewidth}
\graphicspath{{./figures/}}

\allowdisplaybreaks

\definecolor{mycolor01}{rgb}{0.00000,0.00000,1.00000}%
\definecolor{mycolor02}{rgb}{0.133,0.545,0.133}
\definecolor{mycolor03}{rgb}{0.50000,0.00000,0.50000}
\definecolor{mycolor05}{rgb}{1.00000,0.83984,0.00000}
\definecolor{mycolor04}{rgb}{0.92969,0.50781,0.92969}
\definecolor{mycolor06}{rgb}{1.00000,0.64453,0.00000}
\definecolor{mycolor07}{rgb}{0.50000,0.50000,0.50000}
\definecolor{mycolor08}{rgb}{1.00000,0.00000,0.00000}
\definecolor{mycolor09}{rgb}{0.2510 ,0.8784, 0.8157}
\definecolor{mycolor10}{rgb}{0.54297,0.00000,0.00000}
\definecolor{mycolor11}{rgb}{0.6445, 0.1641,0.1641}
\definecolor{mycolor12}{rgb}{1, 0, 1}

\definecolor{colA}{rgb}{0,0,1}%
\definecolor{colB}{rgb}{0.8,0.0,0.5}%
\definecolor{colC}{rgb}{0,0.5,0}%
\definecolor{col_cyan}{rgb}{0,1,1}%
\definecolor{colE}{rgb}{1.00000,0.55,0.00000}%
\definecolor{colF}{rgb}{1.00000,0.0,0.00000}%
\definecolor{col_mag}{rgb}{1.00000,0.00000,1.00000}%
\definecolor{col_grey}{rgb}{0.4,0.4,0.4}%
\definecolor{col_gray2}{rgb}{0.4,0.4,0.4}
\definecolor{lightgray}{rgb}{0.9,0.9,0.9}  

\definecolor{FGgreen}{RGB}{34,139,34}
\definecolor{FGblue}{RGB}{80,120,255}
\definecolor{FGred}{RGB}{255,110,110}

\definecolor{col_agent}{rgb}{1.00000,0.000,1.00000}%
\definecolor{col_bs}{rgb}{0, 0, 0}%
\definecolor{mycolor2}{rgb}{1.00000,0.83984,0.00000}%

\definecolor{col_SP}{RGB}{255,0,0}
\definecolor{col_50k}{RGB}{0,0,255}
\definecolor{col_100k}{RGB}{51, 187, 238}
\definecolor{col_200k}{RGB}{0, 153, 136}

\definecolor{col_1path}{RGB}{34,139,34}
\definecolor{mycolor1}{rgb}{1.00000,0.64453,0.00000}
\definecolor{col_grey}{rgb}{0.4,0.4,0.4}%
\definecolor{col_gray2}{rgb}{0.4,0.4,0.4}
\definecolor{lightgray}{rgb}{0.9,0.9,0.9}  
\definecolor{col_2path}{rgb}{0, 0, 1}
 
 \pgfdeclarelayer{back1}
 \pgfdeclarelayer{lay1}
 \pgfdeclarelayer{back2}
 \pgfdeclarelayer{lay2}
 \pgfsetlayers{back2,lay2,back1,main,lay1}

\makeatletter
 \pgfplotsset{
 	compat=newest,
 	result style group/.style={
 		label style={font=\scriptsize}, 
 		legend style={font=\scriptsize},
 		tick label style={font=\scriptsize},
 		nodes near coords style={font=\scriptsize},
 		title style={font=\scriptsize},
 		axis line style={draw=none},
 		scale only axis,
 		xmajorgrids,
 		ymajorgrids,
 		axis background/.style={fill=white},
 	}
 } 
\tikzset{
  nomorepostactions/.code={\let\tikz@postactions=\pgfutil@empty},
  decmark/.style 2 args={decoration={markings, pre length=#2,
    mark= between positions 0 and 1 step (1/6)*\pgfdecoratedpathlength with{%
        \tikzset{solid, every mark, line width=0.5pt}\tikz@options
        \pgftransformresetnontranslations
        \pgfuseplotmark{#1}%
      },  
    },
    postaction={decorate},
    /pgfplots/legend image post style={
        mark=#1, mark options={solid}, every path/.append style={nomorepostactions}
    },
  },
  posmark/.style 2 args={decoration={markings,
		mark= at position #2*\pgfdecoratedpathlength with{%
			\tikzset{solid,every mark, line width=0.5pt}\tikz@options
			\pgftransformresetnontranslations
			\pgfuseplotmark{#1}%
		},  
	},
	postaction={decorate},
	/pgfplots/legend image post style={
		mark=#1,mark options={solid},every path/.append style={nomorepostactions}
	},
  },
markbeginend/.style 2 args={decoration={markings,
		mark= between positions 0 and 1 step (1)*\pgfdecoratedpathlength with{%
			\tikzset{#2,every mark}\tikz@options
			\pgfuseplotmark{#1}%
		},  
	},
	postaction={decorate},
	/pgfplots/legend image post style={
		mark=#1,mark options={#2},every path/.append style={nomorepostactions}
	},
},
markend/.style 2 args={decoration={markings,
		mark= at position \pgfdecoratedpathlength with{%
			\tikzset{#2,every mark}\tikz@options
			\pgfuseplotmark{#1}%
		},  
	},
	postaction={decorate},
	/pgfplots/legend image post style={
		mark=#1,mark options={#2},every path/.append style={nomorepostactions}
	},
},
}

\makeatother

\pgfplotsset{
resultStyle1/.style={mark=none, line width=0.5pt, mycolor01, decmark={oplus}{0}},
resultStyle2/.style={mark=none, line width=0.5pt, mycolor02, decmark={triangle}{0}},
resultStyle3/.style={mark=none ,line width=0.5pt, mycolor03, decmark={+}{0}},
resultStyle4/.style={mark=none, line width=0.5pt, mycolor06, decmark={star}{0}},
resultStyle5/.style={mark=none, line width=0.5pt, mycolor08, decmark={o}{0}},
resultStyle6/.style={mark=none, line width=0.5pt, mycolor05, decmark={square}{0}}, 
resultStyle7/.style={mark=none, line width=0.5pt, mycolor09, decmark={diamond}{0}}, 
resultStyle8/.style={mark=none, line width=0.5pt, mycolor11, decmark={otimes}{0}}, 
resultStyle9/.style={mark=none, line width=0.5pt, mycolor12, decmark={x}{0}}, 
resultStyleBase/.style={mark=none, line width=0.5pt,}, 
compareStyle1/.style={mark=none, line width=0.5pt, mycolor01},
compareStyle2/.style={mark=none, line width=0.5pt, mycolor02},
compareStyle3/.style={mark=none ,line width=0.5pt, mycolor03},
compareStyle4/.style={mark=none, line width=0.5pt, mycolor06},
compareStyle5/.style={mark=none, line width=0.5pt, mycolor08},
compareStyle6/.style={mark=none, line width=0.5pt, mycolor05}, 
compareStyle7/.style={mark=none, line width=0.5pt, mycolor09}, 
compareStyle8/.style={mark=none, line width=0.5pt, mycolor11}, 
compareStyle9/.style={mark=none, line width=0.5pt, mycolor12}, 
}
  
  \pgfplotsset{
        compat=newest,
        simple style group/.style={
                label style={font=\scriptsize},
                legend style={font=\scriptsize},
                tick label style={font=\scriptsize},
                nodes near coords style={font=\scriptsize},
                title style={font=\scriptsize},
                scale only axis,
                grid style={dotted},
                mark options={solid}, 
        },
        simple style/.style={
                label style={font=\scriptsize},
                legend style={font=\scriptsize},
                tick label style={font=\scriptsize},
                nodes near coords style={font=\scriptsize},
                title style={font=\scriptsize},
                width=\figurewidth,
                height=\figureheight,
                at={(0\figurewidth,0\figureheight)},
                scale only axis,
                grid style={dotted},
                mark options={solid}, 
        },
        base style/.style={
                label style={font=\scriptsize},
                legend style={font=\scriptsize},
                tick label style={font=\scriptsize},
                nodes near coords style={font=\scriptsize},
                title style={font=\scriptsize},
                width=\figurewidth,
                height=\figureheight,
                at={(0\figurewidth,0\figureheight)},
                scale only axis,
                cycle list={
                {mark=none, line width=0.5pt, mycolor01, solid},
                {mark=none, line width=0.5pt, mycolor02, dash dot},
                {mark=none ,line width=0.5pt, mycolor03, densely dashed},
                {mark=none, line width=0.5pt, mycolor04, dash dot dot},
                {mark=x   , line width=0.5pt, mycolor05},
                {mark=.   , line width=0.7pt, mycolor06}, 
                {mark=square,only marks, mark size = 0.8pt, mycolor07,
                mark options = {line width = 0.4pt}},
                {mark=x,     only marks, mark size = 1.3pt, mycolor08,
                mark options = {line width = 0.4pt}},
                {mark=o,     only marks, mark size = 0.8pt, mycolor09,
                mark options = {line width = 0.4pt}},
                {mark=o, mycolor10},
                },
                grid style={dotted},
                xmajorgrids,
                ymajorgrids,
                mark options={solid}, 
        },
        base style group/.style={
        	label style={font=\scriptsize},
        	legend style={font=\scriptsize},
        	tick label style={font=\scriptsize},
        	nodes near coords style={font=\scriptsize},
        	title style={font=\scriptsize},
        	scale only axis,
        	grid style={dotted},
        	xmajorgrids,
        	ymajorgrids,
        	mark options={solid}, 
        },
        std graph style new/.style={
                xlabel style={yshift=1mm},
                ylabel style={yshift=-1.5mm},
                yticklabel style={xshift=1mm},
        },
        color lines style/.style={
                cycle list={
                    {mark=none, mycolor01, decmark={oplus}{0} },
                    {mark=none, mycolor02, decmark={+}{0} }, 
                    {mark=none, mycolor03, decmark={triangle}{0} }, 
                    {mark=none, mycolor04, decmark={star}{0} }, 
                    {mark=none, mycolor05, decmark={o}{0} },
                    {mark=none, mycolor06, decmark={square}{0} },
                },
        },
        meas graph style/.style={
                xlabel style={yshift=1mm},
                ylabel style={yshift=-1mm},
                xmajorgrids,
                ymajorgrids,
                mark repeat = 1,
                mark phase = 0,
                cycle list={
                    {color=black, only marks, mark=*, mark size=0.5pt, mark options={solid, black}},
                    {color=red, only marks, mark=*, mark size=0.1pt, line width=0.25pt},
                },
                ylabel={},
        }, 
        ci graph style/.style={
                xlabel style={yshift=1mm},
                ylabel style={yshift=-1.5mm},
                yticklabel style={xshift=1mm},
                mark repeat = 1,
                mark phase = 0,
                ymin=1e-3,
                ymax=100,
                ytick = {100, 50, 10, 1, 0.1, 0.01, 1e-3, 1e-4},
                yticklabels = {$0$, $50$, $90$, $99$, $99.9$, $99.99$, $99.999$, $99.9999$},
                y dir=reverse,
        },     
        bp coeff style/.style={
               scale only axis=true,
               width=0.225*.9\linewidth,
               height=0.225*.9\linewidth,
               scale only axis,
               xmin=-4.000,
               xmax=4.000,
               xlabel={$\ell${\color{white}$\aod$}},
               ticklabel style={font=\footnotesize},
               ymin=0.000, ymax=0.9,
               ylabel={$c_\ell$},
               xlabel style={font=\footnotesize},
               ylabel style={font=\footnotesize},
               major tick length=2pt
        },
        bp graph style/.style={        
               scale only axis=true,
               width=0.35*1.1\linewidth,
               height=0.225*.9\linewidth,
               scale only axis,
               xmin=-3.14, xmax=3.14,
               xlabel={$\aod${\color{white}$\ell$}},
               ticklabel style={font=\footnotesize},
               xtick={-3.14,-1.57,0.0,1.57,3.14},
               xticklabels={$-\pi$,$-\tfrac{\pi}{2}$,$0$,$\tfrac{\pi}{2}$,$\pi$},
               ymin=0.000, ymax=3,
               ylabel={Beampattern},
               xlabel style={font=\footnotesize}, ylabel style={font=\footnotesize},
               major tick length=2pt
        },
        peb graph style/.style={        
               width=0.66\linewidth,
               scale only axis,
               point meta min=-2.583,
               point meta max=-0.300,
               axis on top,
               xmin=0.000,
               xmax=12.000,
               xlabel={x in meter},
               y dir=reverse,
               ymin=0.000,
               ymax=8.000,
               ylabel={y in meter},
               ytick={7.0,6.0,...,0.0},
               xtick={0.0,1.0,...,12.0},
               yticklabels={$1$,$2$,$3$,$4$,$5$,$6$,$7$,$8$},
               xlabel style={font=\scriptsize,yshift=0.125cm},
               ylabel style={font=\scriptsize,yshift=-0.125cm},
               ticklabel style={font=\scriptsize},
               unit vector ratio*=1 1 1,
               yticklabel pos=left,
               major tick length=2pt,
               colormap={mymap}{[1pt] rgb(0pt)=(1,1,1); rgb(1pt)=(0.858903,0.984776,0.839302); rgb(2pt)=(0.777958,0.94143,0.649487); rgb(3pt)=(0.755504,0.864264,0.463393); rgb(4pt)=(0.777509,0.754439,0.310168); rgb(5pt)=(0.820314,0.619497,0.21003); rgb(6pt)=(0.854796,0.471879,0.170327); rgb(7pt)=(0.851327,0.326629,0.183322); rgb(8pt)=(0.784671,0.198575,0.225774); rgb(9pt)=(0.637629,0.0993149,0.259577); rgb(10pt)=(0.400067,0.0343393,0.229819); rgb(11pt)=(0,0,0)},
               colorbar style={ylabel={Position Error Bound in centimeter (logscale)}, ytick={-0.4,-0.82,...,-2.92}, yticklabels={$39.8$, $15.1$, $5.8$, $2.2$, $0.8$, $0.3$},ylabel style={yshift=0.5mm,font=\scriptsize,scale=0.8},width=2.0mm,xshift=-4.25mm,ticklabel style={font=\scriptsize},major tick length=0pt}, 
               colormap access=piecewise constant
        },
        peb ellipses/.style={color=white, line width=0.4pt, forget plot}
    }

\tikzset{naming/.style={align=center,font=\small}}
\tikzset{antenna/.style={insert path={-- coordinate (ant#1) ++(0,0.25) -- +(135:0.25) + (0,0) -- +(45:0.25)}}}
\tikzset{station/.style={naming,draw,shape=dart,shape border rotate=90, minimum width=10mm, minimum height=10mm,outer sep=0pt,inner sep=3pt}}
\tikzset{mobile/.style={naming,draw,shape=rectangle,minimum width=12mm,minimum height=6mm, outer sep=0pt,inner sep=3pt}}
\tikzset{radiation/.style={{decorate,decoration={expanding waves,angle=90,segment length=4pt}}}}

\tikzset{
  pobl/.style={
    inner sep=0pt, outer sep=0pt, fill=#1,
  },
  pobl gron/.style n args={2}{
    pobl=#1, rounded corners=#2,
  },
  pics/person/.style n args={3}{
    code={
      \node (-corff) [pobl=#1, minimum width=.25*#2, minimum height=.375*#2, rotate=#3, pic actions] {};
      \node (-pen) [minimum width=.3*#2, circle, pobl=#1, outer sep=.01*#2, anchor=south, rotate=#3, pic actions] at (-corff.north) {};
      \node (-coes dde) [pobl gron={#1}{1pt}, anchor=north west, minimum width=.12125*#2, minimum height=.25*#2, rotate=#3, pic actions] at (-corff.south west) {};
      \node [pobl=#1, anchor=north, minimum width=.12125*#2, minimum height=.15*#2, rotate=#3, pic actions] at (-coes dde.north) {};
      \node (-coes chwith) [pobl gron={#1}{1pt}, anchor=north east, minimum width=.12125*#2, minimum height=.25*#2, rotate=#3, pic actions] at (-corff.south east) {};
      \node [pobl=#1, anchor=north, minimum width=.12125*#2, minimum height=.15*#2, rotate=#3, pic actions] at (-coes chwith.north) {};
      \node (-braich dde) [pobl gron={#1}{.75pt}, minimum width=.075*#2, minimum height=.325*#2, outer sep=.0064*#2, anchor=north west, rotate=#3, pic actions] at (-corff.north east)  {};
      \node [pobl=#1, minimum width=.05*#2, minimum height=.2*#2, outer sep=.0064*#2, anchor=north west, rotate=#3, pic actions] at (-corff.north east) {};
      \node (-braich chwith) [pobl gron={#1}{.75pt}, minimum width=.075*#2, minimum height=.325*#2, outer sep=.0064*#2, anchor=north east, rotate=#3, pic actions] at (-corff.north west) {};
      \node [pobl=#1, minimum width=.0375*#2, minimum height=.2*#2, outer sep=.0064*#2, anchor=north east, rotate=#3, pic actions] at (-corff.north west) {};
      \node (-fit person) [fit={(-pen.north) (-braich dde.east) (-coes chwith.south) (-braich chwith.west)}] {};
    },
  },
  pics/SBS/.style={code={
      \begin{scope}[local bounding box=#1]
      \fill [pic actions/.try] (-1,0) -- (-1/2,3) -- (1/2, 3) -- (1,0) -- cycle;
      \fill [pic actions/.try] (-1/16,2) rectangle (1/16,4);
      \fill [pic actions/.try] (0,4) circle [radius=1/4];
      \foreach \i in {-1,1}
        \fill [shift=(90:4), xscale=\i]
          \foreach \r in {1,3/2,2}{
            (-45:\r) arc (-45:45:\r) -- (45:\r-1/10)
            arc(45:-45:\r-1/10) -- cycle
          };
       \end{scope}
  }},
}

\begin{document}

\title{\huge A Sigma Point-based Low Complexity Algorithm for Multipath-based SLAM in MIMO Systems \\[-2mm]
\thanks{
This project was funded by the Christian Doppler Research Association.
} 
}	
\author{
	\IEEEauthorblockN{Anna Masiero, Alexander Venus, and Erik Leitinger}
	\IEEEauthorblockA{\small Graz University of Technology, Graz, Austria,  \{a.masiero, a.venus, erik.leitinger\}@tugraz.at}\\[-10mm]	
}
\maketitle
\frenchspacing


\begin{abstract}
	\Ac{mpslam} is a promising approach in wireless networks to jointly obtain position information of transmitters/receivers and information of the propagation environment. \Ac{mpslam} models specular reflections at flat surfaces as \acp{va}, which are mirror images of base stations. Particle-based methods offer high flexibility and can approximate posterior probability density functions of the mobile agent state and the map feature states, (i.e., \ac{va} states) with complex shapes. However, they often require a large number of particles to counteract degeneracy in high-dimensional parameter spaces, leading to high computational complexity. Conversely using an insufficient number of particles leads to reduced estimation accuracy. 

In this paper, we introduce a low-complexity \ac{mpslam} algorithm using a \ac{sp}-based implementation of the \ac{spa}. We model the messages of continuous states of the agent and the \acp{va} as Gaussian distributions and approximate nonlinearities via \ac{sp}-transformations. This approach substantially reduces the computational complexity without decreasing accuracy. Since probabilistic data association yields Gaussian mixtures for the agent and VA states, we use moment matching to combine each mixture into a single Gaussian. Numerical results using synthetic and real data demonstrate that our method achieves significantly reduced computational runtimes compared to particle-based schemes, while exhibiting comparable (or even superior) localization and mapping performance. 


\end{abstract}

\acresetall 


%
\IEEEpeerreviewmaketitle


\section{Introduction}\label{sec:introduction}
\begin{figure}[t]
	\centering
	\setlength{\abovecaptionskip}{2mm}
	\setlength{\belowcaptionskip}{0pt}
	
	\setlength{\figurewidth}{0.38\textwidth}
	\setlength{\figureheight}{0.1\textwidth}
	\includegraphics[width=.47\textwidth]{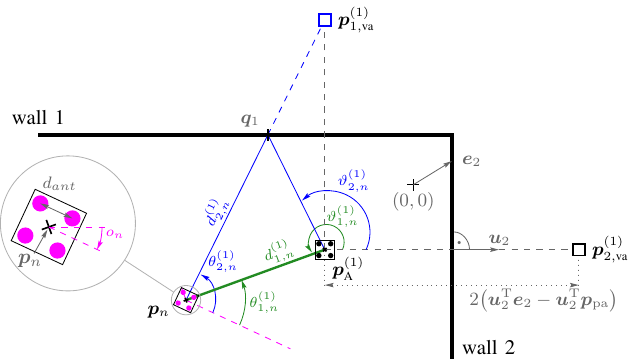}
	\caption{Exemplary indoor environment including the mobile agent at position $\V{p}_n$, a \ac{pa} at position \smash{$\V{p}^{(1)}_\text{A} $} and two corresponding \acp{va} at position $\V{p}_{1,\text{va}}^{(1)}$ and $\V{p}_{2,\text{va}}^{(1)}$. The visualization includes the array geometry used by the agent and \acp{pa}, along with the geometric relationships between the objects. 
	}\label{fig:eye_catcher}
	\vspace{-4mm}
\end{figure}

Emerging sensing and signal processing techniques that exploit multipath propagation promise advanced capabilities in autonomous navigation, asset localization, and situational awareness for future communication networks. \Ac{mpslam} effectively tracks mobile transmitters or receivers while mapping the environment in wireless systems by modelling specular reflections of RF signals as \acp{va}, which are mirror images of \acp{bs} or static transceivers called \acp{pa} (see Fig.~\ref{fig:eye_catcher}) \cite{WitMeiLeiSheGusTufHanDarMolConWin:J16, GentnerTWC2016,KimGraSveKimWym:TVT2022,LeiMeyHlaWitTufWin:J19,LeiVenTeaMey:TSP2023}.

\ac{mpslam} falls under the umbrella of feature-based SLAM approaches, which focus on detecting and mapping distinct environmental features \cite{MonThrKolWeg:AAAI2002, DurrantWhyte2006}. \ac{mpslam} facilitates a \ac{fg}-based representation of the joint posterior density and uses the \ac{spa} to solve the \ac{mpslam} problem in a Bayesian manner. It allows to solve the \ac{pda} problem inherent to \ac{mpslam} with high scalability and was shown to offer a superior trade-off between robustness and runtime \cite{KimGraSveKimWym:TVT2022, LeiMeyHlaWitTufWin:J19, LeiGreWit:ICC2019, MenMeyBauWin:J19}. \ac{mpslam} has been successfully applied to a variety of different scenarios, including cooperative localization \cite{KimGraGaoBatKimWym:TWC2020}, the use of adaptive map feature models \cite{LiLeiCaiTuf:ICC2024}, and environments that involve reflections from rough surfaces \cite{WieVenWilWitLei:Fusion2024}. Most \ac{mpslam} methods use particle-based implementations \cite{AruMasGorCla:TSP2002} to represent the joint posterior distribution  \cite{LeiMeyHlaWitTufWin:J19, LeiGreWit:ICC2019, MenMeyBauWin:J19, KimGraSveKimWym:TVT2022}. Particle-based methods offer high flexibility and can provide an asymptotically optimal approximation of posterior \acp{pdf} with complex shapes. This property is particularly useful for highly nonlinear and reduced information scenarios, such as \ac{toa}-only \ac{mpslam}, where the inherent physics of the problem can induce strongly non-Gaussian \acp{pdf} \cite{LeiMeyHlaWitTufWin:J19}. While the factor graph-based approach to \ac{mpslam} allows for significant reduction of the problem complexity, it typically still requires a high number of particles to counteract particle degeneracy in high-dimensional parameter spaces, leading to high runtimes; conversely using too few particles leads to reduced estimation accuracy. In \ac{mimo} systems, array measurements enable jointly estimating  \ac{toa}, \ac{aoa} and \ac{aod}. The additional information contained in \ac{aoa} and \ac{aod} estimates can yield unambiguous measurement transformations, allowing the resulting joint posterior \ac{pdf} to be approximated accurately by Gaussian densities. A popular method for approximating \acp{pdf} that arise from nonlinear transformations is the unscented or \ac{sp} transform \cite{Julier2004, ArasaratnamHaykin2009_Cubature, Meyer2014_SigmaPointBP}, which has been shown to offer superior approximation performance compared to first-order Taylor linearization employed by \ac{kf}-type methods.

In this paper, we propose an \ac{sp}-based implementation of the \ac{spa} algorithm for \ac{mpslam}. By approximating all \acp{pdf} using \acp{sp}, we efficiently evaluate the integrals required by the algorithm. We describe in detail the steps involved in this approximation, emphasizing the handling of nonlinearities in both the state transition and the measurement models, and discuss the use of moment matching to approximate Gaussian mixtures arising from data association. The main contributions of this paper are as follows.
\begin{itemize}
	\item We propose a novel \ac{sp}-based implementation of the \ac{spa} for \ac{mpslam} leveraging Gaussian approximations of all \acp{pdf} by means of \acp{sp}. This approach allows the integrals required by the algorithm to be evaluated very efficiently.
	\item We provide a detailed derivation of all approximated messages of the \ac{spa}.
	\item We validate our method using simulated data, demonstrating significantly lower runtimes compared to a particle-based implementation, as well as improved accuracy in certain cases.
	\item Using real radio signals, we demonstrate the applicability of the proposed method to real-world scenarios.
\end{itemize}

\textit{Notations and Definitions:} column vectors and matrices are denoted by boldface lowercase and uppercase letters. Random variables are displayed in san serif, upright font, e.g., $\rv{x}$ and $\RV{x}$ and their realizations in serif, italic font, e.g. $x$. $f({x})$ and $p({x})$ denote, respectively, the \ac{pdf} or \ac{pmf} of a continuous or discrete random variable $\rv{x}$ (these are short notations for $f_\rv{x}({x})$ or $p_\rv{x}({x})$). $(\cdot)^{\mathrm{T}}$, denotes the matrix transpose. 
$ \norm{\cdot}{} $ is the Euclidean norm. $ \vert\cdot\vert $ represents the cardinality of a set. $\bdiag\{\V{A}, \V{B}\}$ denotes a block-diagonal matrix with $ \V{A} $ and $ \V{B} $ on the  diagonal and zero matrices in the off-diagonal blocks. $\M{I}_{[\cdot]}$ is an identity matrix of dimension given in the subscript. Furthermore, ${1}_{\mathbb{A}}(\V{x})$ denotes the indicator function that is ${1}_{\mathbb{A}}(\V{x}) = 1$ if $\V{x} \in \mathbb{A}$ and 0 otherwise, for $\mathbb{A}$ being an arbitrary set and $\mathbb{R}^{\text{+}}$ is the set of positive real numbers. The Gaussian \ac{pdf} is $f_\text{N}(x; \hat{x}, \sigma) = 1/(\sqrt{2\pi} \sigma) \exp{(-(x-\hat{x})^2/(2\sigma^2))}$ with mean $\mu$, standard deviation $\sigma$ and the uniform \ac{pdf} $f_\mathrm{U}(x;a,b) = 1/(b-a) {1}_{[a,b]}(x)$. A selected list of acronyms is given in Table~\ref{tab:acro}.

\begin{table}[h]
	\centering
	\footnotesize
	\caption{Selection of acronyms}
	\label{tab:acro}
	\begin{tabular}{|r|l|}
		\toprule
		CEDA    & channel estimation and detection algorithm            \\
		FG      & factor graph                                          \\
		LHF     & likelihood function                                   \\
		MPC     & multipath component                                   \\
		PA      & physical anchor                                       \\
		PVA     & potential virtual anchor                              \\
		RV      & random variable                                       \\
		SP      & sigma point                                           \\
		SPA     & sum-product algorithm                                 \\
		VA      & virtual anchor     \\
		\bottomrule                                  
	\end{tabular}
\end{table}

\section{System Setup and Geometrical Relations}\label{sec:geometric_model}
At each time step $n$, we consider a mobile agent located at position $\bm{p}_{n} \triangleq [{p}_{\text{x}\s n}\;  {p}_{\text{y}\s n}]^\text{T}$, equipped with $\sqrt{N_{\text{ant}}}$ times $\sqrt{N_{\text{ant}}}$ uniform planar array (UPA) with $N_{\text{ant}}$ antenna elements spaced distance $d_{\text{ant}}$ apart and oriented at angle $\kappa_n$. Similarly, $J$ \acp{bs} acting as \acp{pa} and placed at fixed positions $\bm{p}_\text{A}^{(j)}\triangleq [{p}_{\text{x\s A}}^{(j)}\;  {p}_{\text{y\s A}}^{(j)}]^\text{T}$ are also equipped with an $N_{\text{ant}}^{(j)}$-element UPA, with spacing $d_{\text{ant}}^{(j)}$. A radio signal $\V{r}_n^{(j)}$ transmitted by the mobile agent at carrier frequency $f_\text{c}$ and with signal-bandwidth $B$ arrives at the receiver via the line-of-sight (LOS) path as well as via \acp{mpc} originating from the reflection of surrounding objects.

\subsubsection{Feature Model}

Reflections caused by flat surfaces are modelled by \acp{va} \cite{MenMeyBauWin:J19,LeiVenTeaMey:TSP2023,WieVenWilWitLei:Fusion2024}, mirroring the position of the physical anchors on the respective surfaces, located at $\V{p}_{k, \mathrm{va}}^{(j)}=\V{p}_{n}+2(\V{u}_k^{\text{T}} \V{e}_k-\V{u}_k^{\text{T}} \V{p}_{\text{A}}^{(j)}) \V{u}_k$ for first-order reflections, with the vector $\bm{e}_k$ pointing from the coordinate origin to the surface $k$ and the unit vector $\V{u}_k$ normal to that same surface. For notational conciseness, \acp{pa} positions will be referred to as $\bm{p}_\text{A}^{(j)} \triangleq 	\V{p}_{1, \mathrm{va}}^{(j)}$. Further, we denote the distance between the agent and any anchor as $d_{k, n}^{(j)} \triangleq d(\V{p}_{n},\V{p}_{k, \mathrm{va}}^{(j)})=||\V{p}_{n}-\V{p}_{k, \mathrm{va}}^{(j)}||$, the \ac{aoa} as $\theta_{k, n}^{(j)} \triangleq \angle{(\V{p}_n,\V{p}_{k, \mathrm{va}}^{(j)})}+\kappa_n= \atantwo{({p}_{\text{y\s A}}^{(j)}-{p}_{\text{y}\s n}, {p}_{\text{x\s A}}^{(j)}-{p}_{\text{x}\s n})} + \kappa_n$  and the \ac{aod} as $\vartheta_{k, n}^{(j)}=\angle{(\V{p}_n,\V{p}_{\text{A}}^{(j)})}=\atantwo{({p}_{\text{y}\s n}-{p}_{\text{y\s A}}^{(j)}, {p}_{\text{x}\s n}-{p}_{\text{x\s A}}^{(j)})}$ for PAs and $\vartheta_{k, n}^{(j)}=\angle{(\V{p}_n, \V{q}_{k, n}^{(j)})}=\atantwo{({p}_{\text{y}\s n}-{p}_{\text{y}\s k, n}^{(j)}, {p}_{\text{x}\s n}-{p}_{\text{x}\s k, n}^{(j)})}$ for \acp{va} (see Fig.~\ref{fig:eye_catcher}). The reflection point $\V{q}_{k, n}^{(j)} \triangleq [{p}_{\text{x}\s k, n}^{(j)}\;  {p}_{\text{y}\s k, n}^{(j)}]^\text{T}$, needed to relate the \ac{aod} to a \ac{va} is given by\vspace*{-2mm}
\begin{align}
	\V{q}_{k, n}^{(j)}=\V{p}_{k, \mathrm{va}}^{(j)}+\frac{(\V{p}_{\text{A}}^{(j)}-\V{p}_{k, \mathrm{va}}^{(j)})^{\mathrm{T}} \V{u}_k}{2(\V{p}_{n}-\V{p}_{k, \mathrm{va}}^{(j)})^{\mathrm{T}} \V{u}_k}(\V{p}_{n}-\V{p}_{k, \mathrm{va}}^{(j)}).\\[-7mm]\nn
\end{align}

\subsubsection{Measurement Extraction} \label{sec:channel_estimation}

For each time $n$ and anchor $j$, a \ac{ceda} \cite{Hansen2014,HanFleRao:TSP2018, GreLeiWitFle:TWC2024, Moederl2025} extracts an unknown number of measurements $m \in \Set{M}_n^{(j)} \triangleq \{1,\,\dots\,,M_n^{(j)}\} $ from a received RF signal vector $\V{r}_{n}^{(j)}$. Each measurement $\V{z}^{(j)}_{m,n} = [\zd\; \zaoa\; \zaod\; \zu]^{\mathrm{T}}$ contains a distance $\zd = [0, d_{\text{max}}]$, \ac{aoa} $\zaoa=[-\pi,\pi]$, \ac{aod} $\zaod=[-\pi,\pi]$ and normalized amplitude $\zu = [\gamma,\infty)$ component, where $d_{\text{max}}$ is the maximum distance and $\gamma$ the detection threshold of the \ac{ceda}. In effect, channel estimation and detection is a compression of the information contained in $\V{r}_{n}^{(j)}$ into the measurement vector ${\V{z}^{(j)}_{n}}=[\V{z}^{(j)\text{T}}_{1,n} \dots \V{z}^{(j)\text{T}}_{\Set{M}_n^{(j)},n}]$. Note that in contrast to related work, such as \cite{LiLeiVenTuf:TWC2022,VenLeiTerMeyWit:TWC2024}, in this work the normalized amplitude $\zu$ is used exclusively to calculate the measurement variances of $\zd$, $\zaoa$, and $\zaod$ according to \cite{LeiGreWit:ICC2019}.

\section{System Model} \label{sec:system_model}
The state of the mobile agent is given as $\RV{x}_n = [\RV{p}_n^\text{T}\;\RV{v}_n^\text{T}\;\rv{\kappa}_n]^\text{T}$, with its position $\RV{p}_n = [\rv{p}_{\text{x},n}\;\rv{p}_{\text{y},n}]^\text{T}$, velocity $\RV{v}_n = [\rv{v}_{\text{x},n}\;\rv{v}_{\text{y},n}]^\text{T}$, and orientation 
$\rv{\kappa}_n$. In line with \cite{LeiMeyHlaWitTufWin:J19, MeyKroWilLauHlaBraWin:J18}, we account for an unknown number of \acp{va} by introducing \acp{pbo} 
$k \in \{1,\dots, K_{n}^{(j)}\}\triangleq \mathcal{K}_n^{(j)}$. The \ac{pbo} states are denoted as  $\RV{y}^{(j)}_{k,n} \triangleq [\RV{\psi}^{(j)}_{k,n}{}^{\mathrm{T}} \;\rv{r}^{(j)}_{k,n}]^{\mathrm{T}}$, where $\RV{\psi}^{(j)}_{k,n}$ represents the \ac{pbo} position and  $\rv{r}^{(j)}_{k,n} \in \{0,1\}$  is an existence variable modeling the existence/nonexistence of \ac{pbo} $k$, i.e., $r^{(j)}_{k,n} = 1$ if the \ac{pbo} exists. Formally, its state is maintained even if \ac{pbo} $k$ is nonexistent, i.e., if $r^{(j)}_{k,n} = 0$. In that case, the position $\RV{\psi}_{k,n}^{(j)}$ is irrelevant. Therefore, all \acp{pdf} defined for \ac{pbo} states, $f(\V{y}_{k,n}^{(j)}) = f(\V{\psi}_{k,n}^{(j)},r_{k,n}^{(j)})$, are of the form  $f(\V{\psi}_{k,n}^{(j)},r_{k,n}^{(j)}=0) = f_{k,n}^{(j)}\,f_{\mathrm{d}}(\V{\psi}_{k,n}^{(j)})$, where $f_{\mathrm{d}}(\V{\psi}_{k,n}^{(j)})$ is an arbitrary ``dummy \ac{pdf},'' and  $f_{k,n}^{(j)} \in [0,1]$ is a constant representing the probability of non-existence \cite{MeyKroWilLauHlaBraWin:J18, LeiMeyHlaWitTufWin:J19}. Note that for $k \in \{2,\dots, K_{n}^{(j)}\}$, the \acp{pbo} have unknown states $\RV{y}^{(j)}_{k,n}$. In contrast, the \ac{pbo} labeled $k=1$ represent the \ac{pa}, whose position $\RV{\psi}^{(j)}_{1,n}$ is assumed to be known. 
All \acp{pbo} states and agent states up to time $n$ are denoted as $\RV{y}_n \!\triangleq [\RV{y}_n^{(1)\text{T}} \rmv\cdots\ist \RV{y}_n^{(J)\text{T}} ]^{\mathrm{T}}\rmv$ and $\RV{y}_{0:n} \!\triangleq [\RV{y}_0^{\text{T}} \rmv\cdots\ist \RV{y}_n^{\text{T}} ]^{\mathrm{T}}\rmv$ and $\RV{x}_{0:n} \!\triangleq [\RV{x}_0^{\text{T}} \rmv\cdots\ist \RV{x}_n^{\text{T}} ]^{\mathrm{T}}\rmv$, respectively.

\subsection{State Evolution} \label{sec:state_transition_model}
The movement of the agent follows a linear model $\RV{x}_n = \bm{A}\RV{x}_{n-1}+\RV{w}_{n}$, where $\RV{w}_{n}$ is zero mean, Gaussian and i.i.d. across $n$, with covariance matrix $\V{C}_\text{x}$, where we denote the associated state-transition distribution as $f(\V{x}_{n}|\V{x}_{n-1})$.
We distinguish between two types of \acp{pbo}, based on their origin:
\begin{enumerate}
	\item Legacy \acp{pbo} $\underline{\RV{y}}_{k,n}^{(j)}$ ($k \rmv\in\rmv \mathcal{K}_{n-1}^{(j)}$) corresponding to \acp{pbo} that existed at the previous time $\underline{\RV{y}}_{k,n-1}^{(j)}$.
	\item New \acp{pbo} $\overline{\RV{y}}_{m,n}^{(j)}$ ($m \rmv\in\rmv \mathcal{M}_{n}^{(j)}$) appearing at the current time $n$ for the first time\cite{MeyKroWilLauHlaBraWin:J18, LeiMeyHlaWitTufWin:J19}. For each measurement $\V{z}^{(j)}_{n}$ at time $n$ a new \ac{pbo} is introduced.
\end{enumerate}
Legacy \acp{pbo} evolve according to the joint state-transition \ac{pdf}
\begin{align}
	\vspace*{-1mm}
	&f(\V{x}_n ,\underline{\V{y}}_n|\V{x}_{n-1},\V{y}_{n-1})\nn\\[-3mm]
	&\hspace*{15mm} = f(\V{x}_{n}|\V{x}_{n-1})\rmv\rmv \prod_{j=1}^J\prod_{k=1}^{K_{n-1}^{(j)}} \rmv\rmv\rmv\rmv f(\underline{\V{y}}_{k,n}^{(j)} | \V{y}_{k, n-1}^{(j)})
	\label{eq:state_transition}\\[-3mm]\nn
\end{align}
where $ f(\underline{\V{y}}_{k,n}^{(j)}|\V{y}_{k,n-1}^{(j)})\rmv\rmv=\rmv\rmv f(\underline{\V{\psi}}_{k,n}^{(j)}, \underline{r}_{k,n}^{(j)}|\V{\psi}_{k,n-1}^{(j)}, r_{k,n-1}^{(j)}) $ is the augmented state-transition \ac{pdf} assuming that the augmented agent state as well as the \ac{pbo} states evolve independently across $ k $, $ n $ and $j$ \cite{MeyKroWilLauHlaBraWin:J18}. 
At time $n$, a \ac{pbo} that existed at time $ n-1 $ either survives with probability $ p_{\mathrm{s}}$ or dies with probability $1-p_{\mathrm{s}}$. In the case it does survive, its state is distributed according to the state-transition \ac{pdf} $ f(\underline{\V{\psi}}_{k,n}^{(j)}|\V{\psi}_{k,n-1}^{(j)})$, leading to
\begin{align}
	\rmv\rmv\rmv\rmv\rmv\rmv\rmv f(\underline{\V{\psi}}_{k,n}^{(j)}, \underline{r}_{k,n}^{(j)}|\V{\psi}_{k,n\text{-}1}^{(j)}, 1) \rmv = \rmv\rmv
	\begin{cases}
			(1-p_{\mathrm{s}})f_{\mathrm{d}}(\underline{\V{\psi}}_{k,n}^{(j)}),  \rmv\rmv \rmv\rmv\rmv 			&\underline{r}_{k,n}^{(j)} = 0\\
			p_{\mathrm{s}}f(\underline{\V{\psi}}_{k,n}^{(j)}|\V{\psi}_{k,n\text{-}1}^{(j)}),	\rmv\rmv \rmv\rmv\rmv  &\underline{r}_{k,n}^{(j)} = 1
		\end{cases}.\rmv\rmv\rmv\rmv\rmv
	\label{eq:state_transition_pdf2}\\[-6mm]\nn
\end{align}
If a \ac{pbo} did not exist at time $ n\rmv\rmv-\rmv\rmv1 $, i.e., $ r_{k,n-1}^{(j)}\rmv\rmv=\rmv\rmv 0 $, it cannot exist at time $ n $ as a legacy \ac{pbo}, meaning
\begin{align}
	f(\underline{\V{\psi}}_{k,n}^{(j)}, \underline{r}_{k,n}^{(j)}|\V{\psi}_{k,n-1}^{(j)}, 0) =
	\begin{cases}
			f_{\mathrm{d}}(\underline{\V{\psi}}_{k,n}^{(j)}), 	&\underline{r}_{k,n}^{(j)} = 0\\
			0, 											&\underline{r}_{k,n}^{(j)} = 1 
		\end{cases}\,.
	\label{eq:state_transition_pdf1}\\[-7mm]\nn
\end{align}
New \acp{pbo} are modeled by a Poisson point process with mean number of new \ac{pbo} ${\mu}_{\mathrm{n}}$ and \ac{pdf} $f_{\mathrm{n}}(\overline{\V{\psi}}_{m,n}^{(j)})$, where ${\mu}_{\mathrm{n}}$ is assumed to be a known constant. Here, $\overline{r}_{m,n}^{(j)} = 1$ indicates that the measurement $\RV{z}_{m,n}^{(j)}$ was generated by a newly detected \ac{pbo}. New \acp{pbo} become legacy \acp{pbo} at time $n+1$. Accordingly, the number of legacy \acp{pbo} is updated as $K_{n}^{(j)} = K_{n-1}^{(j)} + M_{n}^{(j)}$. To prevent the indefinite growth in the number of \acp{pbo}, \ac{pbo} states with low existence probability (but not \acp{pa}) are removed, as described in Sec. ~\ref{sec:object}.
\subsection{Measurement Model} \label{sec:measurement_model}
Prior to being observed, measurements ${\RV{z}^{(j)}_{n}}$, and consequently their number $\rv{M}^{(j)}_n$, are considered random and are represented by the vector ${\RV{z}^{(j)}_{n}}=[\RV{z}^{(j)\text{T}}_{1,n} \dots \RV{z}^{(j)\text{T}}_{\rv{M}_n^{(j)},n}]$.
Both quantities are stacked into matrices containing all current measurements $\RV{z}_n = [\RV{z}^{(1)\, \text{T}}_{n} ... \, \RV{z}^{(J)\,\text{T}}_{n}]^\text{T}$ and their numbers $\RV{M}_n = [\rv{M}^{(1)}_n \dots \rv{M}^{(J)}_n]$. We assume the \ac{lhf} of a measurement $f(\V{z}_{m,n}^{(j)}| \V{x}_n, \V{\psi}_{k,n}^{(j)})$ to be conditionally independent across its components $\zd$, $\zaoa$ and $\zaod$, i.e.,
\begin{align}
	\rmv f(\V{z}_{m,n}^{(j)}| \V{x}_n, \V{\psi}_{k,n}^{(j)}) \rmv  &= \rmv  f( \zd |\V{p}_{n},\V{p}_{k, \mathrm{va}}^{(j)}) \ist f(\zaod|\V{p}_{n},\V{p}_{k, \mathrm{va}}^{(j)})\nn\\
	&\hspace*{5mm}\times f( \zaoa |\V{p}_{n},\kappa_n, \V{p}_{k, \mathrm{va}}^{(j)}) 
	\label{eq:single_measurement_likelihood}\\[-7mm]\nn
\end{align}
where all factors are given by Gaussian \acp{pdf} (details can be found in \cite{LeiWieVenAsilomar2024_CoopSLAM}). False alarm measurements are assumed to be statistically independent of \ac{pbo} states and are modeled by a Poisson point process with mean $ {{\mu}_{\mathrm{fa}}} $ and \ac{pdf} $ f_{\mathrm{fa}}(\V{z}_{m,n}^{(j)}) $, which is assumed to factorize as $ f_{\mathrm{fa}}(\V{z}_{m,n}^{(j)}) = f_{\mathrm{fa}}(\zd) f_{\mathrm{fa}}(\zaoa)f_{\mathrm{fa}}(\zaod)$. All individual false alarm \acp{lhf} are uniformly distributed in their respective domain. We approximate the mean number of false alarms as $\mu_\text{fa} = N_\text{s}\, e^{-\gamma^2}$, where the right-hand side expression corresponds to the false alarm probability $p_\text{fa}(u) = \int \rmv f_\text{TRayl}(u \,; \sqrt{1/2} \,, \gamma) \, \mathrm{d} u  = e^{-\gamma^2}$ \cite[p. 5]{VenLeiTerMeyWit:TWC2024}.
\subsection{Data Association Uncertainty}
\label{sec:DA}

The inference problem at hand is complicated by the data association uncertainty: at time $n$, it is unknown which measurement $\V{z}_{m,n}^{(j)}$ (extracted with detection probability $p_{\text{d}}$ from \ac{pa} $j$) originates from a \ac{pbo}, a \ac{pa}, or clutter. Moreover, one has take into account missed detections and the possibility that a \ac{pbo} has just become visible or obstructed during the current time step $n$.
In line with \cite{LeiMeyHlaWitTufWin:J19,MeyKroWilLauHlaBraWin:J18}, we apply the ''point object assumption'', i.e. we assume that each \ac{pbo} generates at most one measurement and each measurement is generated by at most one \ac{pbo}, per time $n$. We use a redundant formulation of the data association problem using two association vectors $ \underline{\RV{a}}_{n}^{(j)} \triangleq [\underline{\rv{a}}_{1,n}^{(j)} \ist \cdots \ist  \underline{\rv{a}}_{\rv{K}_{n-1},n}^{(j)}]^{\mathrm{T}} $ and $ \overline{\RV{a}}_{n}^{(j)} \triangleq [\overline{\rv{a}}_{1,n}^{(j)} \ist \cdots \ist \overline{\rv{a}}_{\rv{M}_{n},n}^{(j)}]^{\mathrm{T}}$ leading to an algorithm that is scalable for large numbers of \acp{pbo} and measurements \cite{WilLau:J14,LeiMeyHlaWitTufWin:J19,MeyKroWilLauHlaBraWin:J18}. The first variable, $\underline{\rv{a}}_{k,n}^{(j)} $ takes values $ m \in \{0, 1, \dots, {M}_n^{(j)}\}$, is \ac{pbo}-oriented indicating which measurement $m$ was generated by \ac{pbo} $k$, where $0$ represents the event that no measurement was generated by \ac{pbo} $k$ (missed detection). The second variable $\overline{\rv{a}}_{m,n}^{(j)}$ is measurement-oriented taking values $ k \in \{0, 1, \dots, {K}_n^{(j)}\}$ and specifying the source $k$ of each measurement $m$, where $0$ represents a measurement not originating from a legacy \ac{pbo} (i.e, it originates from a new  \ac{pbo} or clutter). To enforce the point target assumption the exclusion functions $ \Psi(\underline{\V{a}}_n^{(j)},\overline{\V{a}}_n^{(j)}) $ and $\Gamma_{\overline{a}_{m,n}^{(j)}}(\overline{r}_{m,n}^{(j)}) $ are applied. The former prevents two legacy \acp{pbo} from being generated by the same measurement, while the latter ensures that a measurement cannot be generated by both a new \ac{pbo} and a legacy \ac{pbo} simultaneously.
The function $ \Psi(\underline{\V{a}}_n^{(j)},\overline{\V{a}}_n^{(j)}) \triangleq \prod_{k = 1}^{K_{n-1}^{(j)}}\prod_{m = 1}^{M_{n}^{(j)}}\psi(\underline{a}_{k,n}^{(j)},\overline{a}_{m,n}^{(j)})$ is defined by its factors, given as
\vspace*{-1mm}
\begin{align}
	\psi(\underline{a}_{k,n}^{(j)},\overline{a}_{m,n}^{(j)}) \rmv\rmv\triangleq \rmv\rmv
\begin{cases}
	0 , & \text{\parbox{\textwidth/4}{$ \underline{a}_{k,n}^{(j)} = m $ and $ \overline{a}_{m,n}^{(j)} \neq k $ or\\[0mm] $ \overline{a}_{m,n}^{(j)} = k $ and $ \underline{a}_{k,n}^{(j)} \neq m $}}\\[1mm]
	1,					     & \text{else}\label{eq:exclusion_psi}  \\[-1mm]
\end{cases}\\[-7mm]\nn
\end{align}
and $ \Gamma_{\overline{a}_{m,n}^{(j)}}(\overline{r}_{m,n}^{(j)})$ is given as
\vspace*{-1mm}
\begin{align}
\Gamma_{\overline{a}_{m,n}^{(j)}}(\overline{r}_{m,n}^{(j)})  \rmv\rmv\triangleq \rmv\rmv
	\begin{cases}
			0 , & \text{\parbox{\textwidth/4}{$\overline{r}_{m,n}^{(j)} = 1$ and $ \overline{a}_{m,n}^{(j)} \neq 0  $}} \\[1mm]
			1,					     &\text{else} \label{eq:exclusion_gamma} \\[-1mm]
		\end{cases}.\\[-7mm]\nn 
\end{align}
The joint vectors containing all association variables for times $n$ are given by $\underline{\RV{a}}_{n} \triangleq [\underline{\RV{a}}_{1}^{(j)\s \mathrm{T}} \, ... \, \underline{\RV{a}}_{n}^{(j)\s\mathrm{T}} ]^{\mathrm{T}}$, $\overline{\RV{a}}_{n} \triangleq [\overline{\RV{a}}_{1}^{(j)\s \mathrm{T}} \, ... \, \overline{\RV{a}}_{n}^{(j)\s \mathrm{T}} ]^{\mathrm{T}}$.


\section{Problem Formulation and Proposed Method} \label{sec:factor_graph}
In this section we formulate the estimation problem, introduce the joint posterior distribution, and outline proposed \acf{spa}.

\subsection{Problem Formulation and State Estimation}\label{sec:object}

We aim to estimate the agent state $\V{x}_n$ considering all measurements $\V{z}_{1:n}$ up to the current time $n$. In particular, we calculate an estimate by using the \ac{mmse}, which is given as \cite{Kay1993} 
\begin{align}\label{eq:mmse} 
	\hat{{\bm{x}}}^\text{MMSE}_{n} \,\triangleq \int \rmv {\bm{x}}_{n} \, f({\bm{x}}_{n} |\V{z}_{1:n} )\, \mathrm{d}{\bm{x}}_{n} 
\end{align}
with $\hat{{\bm{x}}}^\text{MMSE}_{n}=[\hat{\V{p}}_n^\text{MMSE T}\;\hat{\V{v}}_n^\text{MMSE T}\;\hat{\kappa}_n^\text{MMSE}]^\text{T}$. We also aim to determine an estimate of the environment map, represented by an unknown number of \acp{pbo} with their respective positions $\bm{\psi}^{(j)}_{k,n}$. To this end, we determine the marginal posterior existence probabilities $p({r}_{k,n}^{(j)} \!=\!1 \big| \V{z}_{1:n}) =\rmv \int \rmv f({\V{\psi}}_{k,n}^{(j)}\ist, r_{k,n}^{(j)} \!=\! 1\big|\V{z}_{1:n}) \ist\mathrm{d}{\V{\psi}}_{k,n}^{(j)}$ and the marginal posterior PDFs $f(\V{\psi}_{k,n}^{(j)} | r_{k,n}^{(j)} = 1,\V{z}_{1:n}) = f(\V{\psi}_{k,n}^{(j)}, r_{k,n}^{(j)} = 1 | \V{z}_{1:n})/ p(r_{k,n}^{(j)} = 1 | \V{z}_{1:n})$. A \ac{pbo} $\V{\psi}_{k,n}^{(j)}$ is declared to exist if $ p(r_{k,n}^{(j)} = 1|\V{z}_{1:n}) > p_\mathrm{de}$, where $p_\mathrm{de}$ is a detection threshold. To avoid that the number of \ac{pbo} states grows indefinitely, \ac{pbo} states with $p(r_{k,n}^{(j)} = 1|\V{z}_{1:n}) < p_{\mathrm{pr}}$ are removed from the state space. For existing \acp{pbo}, a position estimate $\bm{\psi}_{k,n}^{(j)}$ is again calculated by the \ac{mmse} \cite{Kay1993} 
\begin{align}\label{eq:mmsepbo}
	\hat{\bm{\psi}}^{(j)\s\text{MMSE}}_{k,n}\triangleq \int \rmv \bm{\psi}_{k,n}^{(j)} \, f(\bm{\psi}_{k,n}^{(j)} | r_{k,n}^{(j)} = 1 , \V{z}_{1:n} )\, \mathrm{d} \bm{\psi}_{k,n}^{(j)}.
\end{align}
As direct computation of marginal distributions from the joint posterior $f( \V{x}_{0:n}, \V{y}_{1:n}, \underline{\V{a}}_{1:n}, \overline{\V{a}}_{1:n} ,\V{m}_{1:n} | \V{z}_{1:n} )$ is infeasible \cite{MeyKroWilLauHlaBraWin:J18}, we perform message passing on the factor graph that represents the factorization of the joint distributions. The messages at issue are computed efficiently by applying a Gaussian approximation to all \acp{pdf}.

\subsection{Joint Posterior and Factor Graph} \label{sec:joint_posterior}
Applying Bayes' rule {as well as some commonly used independence assumptions}\cite{MeyKroWilLauHlaBraWin:J18,LeiMeyHlaWitTufWin:J19}, the joint posterior \ac{pdf} is given as
\vspace*{-1mm}
\begin{align}
	&f( \V{x}_{0:n}, \V{y}_{1:n}, \underline{\V{a}}_{1:n}, \overline{\V{a}}_{1:n} ,\V{m}_{1:n} | \V{z}_{1:n} ) \nn \\
	&\hspace{4mm}\propto  (f(\V{x}_{0}) \prod^{J}_{j'=1}\rmv\rmv\rmv f(\underline{\V{y}}^{(j')}_{1,0}))
	\prod^{n}_{n'=1} \rmv \! \Phi_\mathrm{x}(\V{x}_{n'}|\V{x}_{n'-1}) \nn\\[-2mm]
	&\hspace{4mm}\times (\prod^{J}_{j=1}\underline{g}\big( \V{x}_{n'}, \underline{r}^{(j)}_{1,n'}, \underline{a}^{(j)}_{1,n'}; \V{z}^{(j)}_{n'} \big) \prod^{M_{n'}}_{m=1} \rmv\rmv\Psi\big(\underline{\V{a}}^{(j)}_{1,n'} \rmv,\overline{\V{a}}^{(j)}_{m,n'}\big))\nn\\
	&\hspace{4mm}\times\prod^{J}_{j=1} \rmv\rmv\Psi\big(\underline{\V{a}}^{(j)}_{n'} \rmv,\overline{\V{a}}^{(j)}_{n'}\big) \prod^{K^{(j)}_{n'-1}}_{k=2}\rmv\rmv\rmv \Phi_k\big(\underline{\V{y}}^{(j)}_{k,n'} \big| \V{y}^{(j)}_{k,n'-1}\big) \nn\\[-1mm]
	&\hspace{4mm}\times \rmv 
	\underline{g}\big( \V{x}_{n'}, \underline{\V{\psi}}^{(j)}_{k,n'} , \underline{r}^{(j)}_{k,n'}, \underline{a}^{(j)}_{k,n'}; \V{z}^{(j)}_{n'} \big)\rmv
	\nn \\[-2mm]
	&\hspace{4mm}\times \prod^{M^{(j)}_{n'}}_{m=1} \overline{g}\big( \V{x}_{n'}, \overline{\V{\psi}}^{(j)}_{m,n'} , \overline{r}^{(j)}_{m,n'}, \overline{a}^{(j)}_{m,n'}; \V{z}^{(j)}_{n'} \big)  
	\label{eq:factorization1}\\[-6mm]\nn
\end{align}
where we introduced the state-transition functions  $\Phi_\mathrm{x}({\bm{x}}_n|{\bm{x}}_{n-1}) \triangleq  f({\bm{x}}_n|{\bm{x}}_{n-1})$, and $\Phi_k(\underline{\bm{y}}_{k,n}^{(j)}|\bm{y}_{k,n-1}^{(j)}) \triangleq  f(\underline{\bm{y}}_{k,n}^{(j)}|\bm{y}_{k,n-1}^{(j)})$, as well as the pseudo \acp{lhf} $\underline{g}\big( \V{x}_n, \underline{\V{\psi}}^{(j)}_{k,n} , \underline{r}^{(j)}_{k,n}, \underline{a}^{(j)}_{k,n}; \V{z}^{(j)}_{n} \big)$ and  $\overline{g}\big( \V{x}_n, \overline{\V{\psi}}^{(j)}_{m,n} , \overline{r}^{(j)}_{m,n}, \overline{a}^{(j)}_{m,n}; \V{z}^{(j)}_{n} \big)$, for legacy \acp{pbo} and new \acp{pbo}, respectively.
\vspace{-0.5mm}
For $\underline{g}\big( \V{x}_n, \underline{\V{\psi}}^{(j)}_{k,n} , \underline{r}^{(j)}_{k,n}, \underline{a}^{(j)}_{k,n}; \V{z}^{(j)}_{n} \big)$ one obtains
\vspace*{-2mm}
\begin{align}
	&\underline{g}\big( \V{x}_n, \underline{\V{\psi}}^{(j)}_{k,n} , 1, \underline{a}^{(j)}_{k,n}; \V{z}^{(j)}_{n} \big)  \nn \\[1.5mm]
	&\hspace{1mm}=\begin{cases}
			\displaystyle \frac{ \pd \ist f(\V{z}_{m,n}^{(j)}| \V{x}_n, \underline{\V{\psi}}_{k,n}^{(j)}) }  {\mu_\mathrm{fa} \ist f_{\mathrm{fa}}\big( \V{z}^{(j)}_{m,n} \big)} \ist, 
			&\rmv \underline{a}^{(j)}_{k,n}\!=\rmv m \in\! \Set{M}^{(j)}_n \\[4.5mm]
			1 \!-\rmv \pd \ist, &\rmv \underline{a}^{(j)}_{k,n} \!=\rmv 0 
		\end{cases}\label{eq:underlineg} \\[-6mm]\nn
\end{align}
and $\underline{g}\big( \V{x}_n, \underline{\V{\psi}}^{(j)}_{k,n} , 0, \underline{a}^{(j)}_{k,n}; \V{z}^{(j)}_{n} \big) = 1_{\{0\}}\big(\underline{a}^{(j)}_{k,n}\big)$. 
Similarly, for $\overline{g}\big( \V{x}_n, \overline{\V{\psi}}^{(j)}_{m,n} , \overline{r}^{(j)}_{m,n}, \overline{a}^{(j)}_{m,n}; \V{z}^{(j)}_{n} \big)$ one can write 
\vspace*{-2mm}
\begin{align}
	&\overline{g}\big( \V{x}_n, \overline{\V{\psi}}^{(j)}_{m,n} , 1, \overline{a}^{(j)}_{m,n}; \V{z}^{(j)}_{n} \big) \nn \\[1mm]
	&\triangleq \begin{cases}
			0 \ist,  &\rmv\rmv\rmv\rmv\rmv\rmv \overline{a}^{(j)}_{m,n} \!\in\rmv\Set{K}^{(j)}_{n-1} \\[1.5mm]
			\displaystyle \frac{\mu_{\mathrm{n}} \ist f_{\mathrm{n}}(\overline{\V{\psi}}_{m,n}^{(j)})  \ist f(\V{z}_{m,n}^{(j)}| \V{x}_n, \overline{\V{\psi}}_{m,n}^{(j)}) }{\mu_\mathrm{fa} \ist f_{\mathrm{fa}}\big( \V{z}^{(j)}_{m,n} \big)} 
			\ist, &\rmv\rmv\rmv\rmv\rmv\rmv \overline{a}^{(j)}_{m,n} \rmv=\rmv 0 
			\label{eq:overlineg}
		\end{cases}
\end{align}
and $\overline{g}\big( \V{x}_n, \overline{\V{\psi}}^{(j)}_{m,n} , 0, \overline{a}^{(j)}_{m,n}; \V{z}^{(j)}_{n} \big) \rmv\triangleq\rmv f_{\mathrm{d}}\big(\overline{\V{\psi}}^{(j)}_{m,n}\big)$. 
A detailed derivation of \eqref{eq:factorization1} is provided in \cite{LeiMeyHlaWitTufWin:J19, VenLeiTerMeyWit:TWC2024}.

\subsection{Sum-Product Algorithm (SPA)}  \label{sec:spa}

\begin{figure*}[!t]
	\vspace{-3.8cm}
	\centering
\begin{subfigure}{.57\textwidth}
	\tikzsetnextfilename{factor_graph_full}
	\scalebox{0.75}{\includegraphics[]{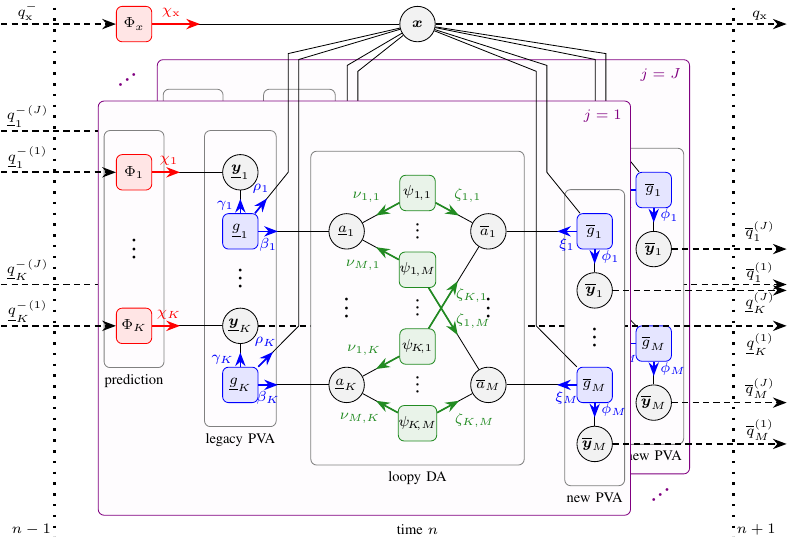}}
	\caption{}
	\label{fig:factorGraph}
\end{subfigure}\hfill
\begin{subfigure}{.38\textwidth}
	\setlength{\abovecaptionskip}{0mm}
	\setlength{\belowcaptionskip}{0pt}
	\tikzsetnextfilename{nlos_floorplan}
	\scalebox{1}{\hspace{-11mm}\includegraphics[]{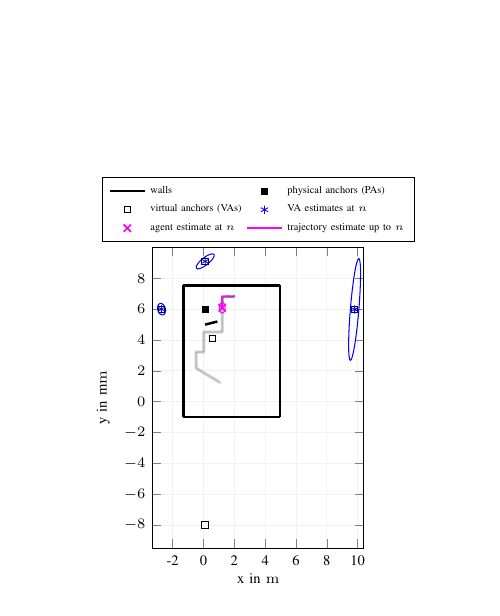}}
	\caption{}
	\label{fig:floorplan}
\end{subfigure}
\vspace*{-1mm}
\caption{(a) Factor graph corresponding to the factorization shown in \eqref{eq:factorization1}. 
	Dashed arrows represent messages that are only passed in one direction. The following short notations are used: 
	$ K \triangleq K_{n-1}^{(j)} $, 
	$ M \triangleq M_{n}^{(j)} $; 
	\emph{variable nodes}: 
	$ \underline{a}_{k} \triangleq \underline{a}_{k,n}^{(j)} $, 
	$ \overline{a}_{m} \triangleq \overline{a}_{m,n}^{(j)} $, 
	$ \V{x}\triangleq \V{x}_{n} $,
	$ \underline{\V{y}}_{k} \triangleq \underline{\V{y}}_{k,n}^{(j)} $, 
	$ \overline{\V{y}}_{m} \triangleq \overline{\V{y}}_{m,n}^{(j)} $;
	\emph{factor nodes}: 
	$ \Phi_{x} \triangleq \Phi_x(\V{x}_{n} | \V{x}_{n} ) $, 
	$ \Phi_{k} \triangleq \Phi_k(\underline{\V{y}}_{k,n}^{(j)} |\V{y}_{k,n-1}^{(j)} ) $, 
	$ \underline{g}_{k} \triangleq \underline{g}( \V{x}_{n}, \underline{\V{\psi}}^{(j)}_{k,n} , \underline{r}^{(j)}_{k,n}, \underline{a}^{(j)}_{k,n}; \V{z}^{(j)}_{n} ) $, 
	$ \overline{g}_{m} \triangleq  \overline{g}( \V{x}_{n}, \overline{\V{\psi}}^{(j)}_{m,n} , \overline{r}^{(j)}_{m,n}, \overline{a}^{(j)}_{m,n}; \V{z}^{(j)}_{n} ) $, 
	$ \psi_{k,m} \triangleq \psi(\underline{a}_{k,n}^{(j)},\overline{a}_{m,n}^{(j)}) $; 
	\emph{prediction}: 
	$ \chi_{k} \triangleq \chi( \underline{\V{\psi}}_{k,n}^{(j)},  \underline{\rv{r}}_{k,n}^{(j)} ) $, 
	$ \chi_\text{x} \triangleq \chi_\text{x}( \V{x}_n)$; 
	\emph{measurement evaluation}: 
	$ \beta_{k} \triangleq \beta(\underline{a}_{k,n}^{(j)}) $, $ \xi_{m} \triangleq \xi(\overline{a}_{m,n}^{(j)}) $; 
	\emph{loopy DA}: 
	$ \nu_{m,k} \triangleq \nu_{m \rightarrow k}(\underline{a}_{k,n}^{(j)}) $, 
	$ \zeta_{k,m} \triangleq \zeta_{k \rightarrow m}(\overline{a}_{m,n}^{(j)}) $, 
	$ \eta_{k} \triangleq \eta(\underline{a}_{k,n}^{(j)}) $, 
	$ \varsigma_{m} \triangleq \varsigma(\overline{a}_{m,n}^{(j)}) $; 
	\emph{measurement update}: 
	$ \gamma_{k} \triangleq \gamma(  \underline{\V{\psi}}_{k,n}^{(j)},  \underline{\rv{r}}_{k,n}^{(j)}  ) $, 
	$ \rho_{k} \triangleq \rho_{k}^{(j)}(  \V{x}_n ) $, 
	$ \phi_{m} \triangleq \phi( \overline{\V{\psi}}^{(j)}_{m,n}, \overline{r}_{m,n}^{(j)}) $, 
	$ \kappa_{m} \triangleq \kappa_{m}^{(j)}(  \V{x}_n ) $; 
	\emph{belief calculation:} 
	$q_\mathrm{x} \triangleq q(\V{x}_{n}), 
	\underline{q}_{k}^{(j)} \triangleq {q}(\underline{\V{y}}_{k,n}^{(j)}),
	\overline{q}_{m}^{(j)} \triangleq	{q}(\overline{\V{y}}_{m,n}^{(j)}),
	q_\mathrm{x}^{-} \triangleq q(\V{x}_{n\minus 1}), 
	\underline{q}_{k}^{- (j)} \triangleq {q}(\underline{\V{y}}_{k,n\minus 1}^{(j)})
   $. (b) Scenario used to generate synthetic data showing the true map, i.e. one PA with its VAs, the room and a wall temporarily obstructing the LOS, as well as the estimated agent and VA positions along with a visualization of their covariance matrix (100-fold) at time $n=52$.}
   \vspace{-4mm}
\end{figure*}

To compute the marginal distributions of Eq.~\eqref{eq:factorization1}, we apply \ac{bp} by means of the \acf{spa} rules \cite{KscFreLoe:TIT2001,Loe:SMP2004_FG} on the \ac{fg} depicted in Fig.~\ref{fig:factorGraph}. A full derivation of these messages and the scheduling used to solve the graph is provided in the supplementary material of \cite{VenLeiTerMeyWit:TWC2024}.


\subsection{Sigma Point Implementation}\label{sec:sp}
Since the message integrals of the proposed \ac{spa} \cite{VenLeiTerMeyWit:TWC2024} corresponding to continuos \acp{rv} cannot be solved analytically, we approximate the according posterior distributions as Gaussian.
The nonlinear measurement model is handled by means of the \ac{sp} transform, which requires calculating a set $\{(\V{s}^{(i)}, w_m^{(i)}, w_c^{(i)}\}_{i=0}^I$ of $I$ points, called \acfp{sp}. The points $\V{s}^{(i)}$ and their corresponding weights $w_m^{(i)}$ and $w_c^{(i)}$ are calculated from a Gaussian \ac{pdf} with mean vector $\V{\mu}_\text{s}$ and covariance  matrix $\V{C}_\text{s}$ according to \cite[Eq. 12]{Julier2004}. The \acp{sp} are then propagated through the nonlinear function $\V{t}^{(i)}=H(\V{s}^{(i)})$, resulting in the set $\{(\V{s}^{(i)},\V{t}^{(i)}, w_m^{(i)}, w_c^{(i)}\}_{i=0}^I$ from which the approximated mean, covariance and cross-covariance are calculated as \cite[Eq. 9-10]{Julier2004}
\vspace*{-1mm}
\begin{align}\label{eq:sp1}
	\hspace*{-2mm}\tilde{\V{\mu}}_\text{t} = \sum_{i=0}^{2I} w_m^{(i)} \V{t}^{(i)}, \iist\iist \tilde{\V{C}}_\text{t} = \sum_{i=0}^{2I} w_c^{(i)} (\V{t}^{(i)}-\tilde{\V{\mu}}_\text{t}) (\V{t}^{(i)}-\tilde{\V{\mu}}_\text{t})\transp \\[-9mm]\nn
\end{align}
\begin{align}\label{eq:sp2}
	\text{and} \quad \tilde{\V{C}}_\text{st} = \sum_{i=0}^{2I} w_c^{(i)} (\V{s}^{(i)}-\V{\mu}_\text{s}) (\V{t}^{(i)}-\tilde{\V{\mu}}_\text{t})\transp.\\[-7mm]\nn
\end{align}
Expressions involving the measurement model (Sec.~\ref{sec:measurement_model}) are approximated by the equations given above as shown explicitly in \cite{Meyer2014_SigmaPointBP}. Independent states can be stacked into a joint state vector, which then requires an only set of \acp{sp} to be represented. Note that since the posterior distributions are approximated as Gaussian, the approximated integrals of the individual \ac{spa} messages take the form of standard \ac{kf} prediction and update equations.
What follows is an overview of the resulting algorithm, with the details of each step displayed in the appendix to this paper.
\begin{enumerate}[]
	\item The prediction messages of agent and legacy PVAs are calculated by applying the \ac{kf} prediction equation to the beliefs of the previous time.
	\item The measurements are evaluated using expressions related to marginal and conditional Gaussian \acp{pdf}, where a SP transform is applied to handle the measurement models nonlinearity. When considering new PVAs, their uniform prior isn't approximated by a Gaussian \ac{pdf}, as that would lead to inaccurate results. Instead, we use importance sampling to represent the uniform distribution.
	\item The results of step 2 are fed into the loopy DA.
	\item Existences are calculated for new and legacy PVAs using results from step 2.
	\item The beliefs are evaluated using the \ac{kf} update equation and \ac{sp} transform with previously obtained results. The evidence term needs to be considered as stated in Bayes' theorem, as \ac{kf} update only provides the posterior distribution. The Gaussian mixture found in the agent update is approximated using moment matching.
\end{enumerate}

\section{Numerical Evaluation}\label{sec:results}
We validate the proposed algorithm in a numerical simulation and compare against the performance of a MIMO implementation of particle-based \ac{mpslam} following \cite{LeiMeyHlaWitTufWin:J19,LeiGreWit:ICC2019, LeiWieVenAsilomar2024_CoopSLAM}, using $1\ist000$, $10\ist000$, $50\ist000$ and $100\ist000$ particles. We further validate the algorithm through a small measurement campaign. For the agent, the positioning error is quantified in terms of the \ac{rmse} and the empirical cumulative distribution function (eCDF) of the error magnitude $e_{\text{pos}}$. Note that the first two steps, i.e. the initialization steps, are not considered for the eCDF plot. For \acp{va} we evaluate the mean optimal subpattern assignment (OSPA) with a cut-off parameter of 5 and order of 2 \cite{SchVoVo:TSP2008} and the cardinality error. Furthermore, we determine the Cram\`er-Rao Lower Bound (CRLB) \cite{TichavskyTSP1998,LeitingerJSAC2015,KalGeTalWymVal:Fusion2021} as a benchmark and the mean runtime of the algorithm per iteration.

\subsection{Simulation and Algorithm Parameters}\vspace{-2.2mm}
The scenario's geometry depicted in Fig. \ref{fig:floorplan} shows the agent's $300$-step long trajectory through an approximately $6.5\mathrm{m}\times 7.5\mathrm{m}$ sized room, equipped with one physical anchor. Measurements are generated according to the model described in Sec.~\ref{sec:measurement_model} only considering first order reflections. The signal is transmitted at $f_c=6\mathrm{GHz}$ with a bandwidth of $B=500\mathrm{MHz}$ and a root-raised cosine pulse shape with roll-off factor $\beta=0.6$. The signal power follows a free-space path loss model and is equal to $40\mathrm{dB}$ at one-meter distance with each reflection causing a $3\mathrm{dB}$ attenuation. Receiver and transmitter both have a $3\times 3$ antenna array, each element spaced $d_\text{ant}=\frac{\lambda}{4}$ apart. The mean number of false alarms is approximated according to $\mu_{\text{fa}} = 2N_{\text{ant}}\cdot \exp{-\gamma^2}$ \cite{VenLeiTerMeyWit:TWC2024}. A detection threshold of $\gamma = 9\mathrm{dB}$ was set, resulting in a mean number of false alarms of $\mu_{\text{fa}}\approx 5$. Experiments were performed with $500$ realizations, except when the particle-based \ac{mpslam} with $100\ist000$ particles was involved, in which case the realizations were reduced to $200$. New \acp{pbo} are initialized with a mean number of $\mu_{\text{n}}=0.1$ and distributed uniformly on a disc with radius $d_{\text{max}}=15\mathrm{m}$. The survival probability is set to $p_s = 0.999$ and the threshold of existence above which a VA is considered detected or lost equals $p_{\text{de}}=0.5$ and $p_{\text{pr}}=10^{-4}$ respectively. The loopy data association performs a maximum of $N_{\text{DA}}=10^5$ message passing iterations and the number of samples $P$, used to approximate the distribution of new \acp{pbo} is $P=10$. Further, we model the movement of the agent according to the continuous velocity and stochastic acceleration model  $\RV{x}_n = \bm{A} \RV{x}_{n-1} + \bm{B} \RV{w}_{n}'$ detailed in \cite[p. ~273]{BarShalom2002EstimationTracking}, where $\RV{w}_{n}'$ is a zero mean Gaussian noise process, i.i.d. across $n$, and with covariance matrix ${\sigma_{\text{a}}^2}\, \bm{I}_2$. Here, ${\sigma_{\text{a}}^2}$ denotes the acceleration standard deviation, and the state transition matrices are given as
\vspace*{-1mm}
\begin{align*}
	\bm{A}  =   \begin{bmatrix}
			1 &  \Delta T \\
			0 &  1\\
		\end{bmatrix}  \otimes \V{I}_{N_\text{D}} \hspace{1cm}\text{and}\hspace{1cm} \bm{B}  =   \begin{bmatrix}
			\frac{\Delta T^2}{2} \\
			\Delta T\\
		\end{bmatrix}  \otimes \V{I}_{N_\text{D}}\\[-6mm]
\end{align*}
with $\Delta T$ as the observation period, set to $1\mathrm{s}$. The model is rewritten to fit the model in Sec.~\ref{sec:state_transition_model} by setting $\RV{w}_{n} \triangleq \bm{B}\RV{w}_{n}'$, where $\RV{w}_{n}$ is still zero mean and i.i.d. across $n$, but with covariance matrix
\vspace*{-1mm}
\begin{align*}
	\boldsymbol{C}_\text{x}=\left[\begin{array}{cc}
			\frac{\Delta T^4}{4} & \frac{\Delta T^3}{2} \\
			\frac{\Delta T^3}{2} & \Delta T^2
		\end{array}\right] \otimes \boldsymbol{I}_{N_{\mathrm{D}}} \sigma_{\mathrm{a}}^2.\\[-6mm]
\end{align*}
The velocity state transition noise is chosen to be $\sigma_{\text{a}}^2 = 9\cdot 10^{-4}\mathrm{m/s^2}$ according to \cite[p. 274]{BarShalom2002EstimationTracking} and the orientation variance to $\sigma_{\text{a}}^2 = 5^\circ$. The initial agent state is drawn from a normal distribution centred around the true agent position with standard deviations $\sigma_{\text{p},0}=0.1\mathrm{m}$, $\sigma_{\text{v},0}=0.01\mathrm{m/s}$ and $\sigma_{\kappa,0}=10^\circ$ for its position, velocity and orientation. The location of all PAs is assumed to be fixed and known. A small regularization noise with variance ${\sigma_{\text{reg}}^2} = 0.01^2\mathrm{m}$ is added to the \ac{pbo} positions for numerical reasons in a pseudo state-transition with covariance matrix $\sigma_{\text{reg}}^2\M{I}_{[2]}$.

\begin{figure*}[t]	
	\centering
	\setlength{\abovecaptionskip}{0mm}
	\setlength{\belowcaptionskip}{0pt}
	
	\setlength{\figurewidth}{0.38\textwidth}
	\setlength{\figureheight}{0.1\textwidth}
	
	\tikzsetnextfilename{mimo_combi}
	\scalebox{1}{\includegraphics[]{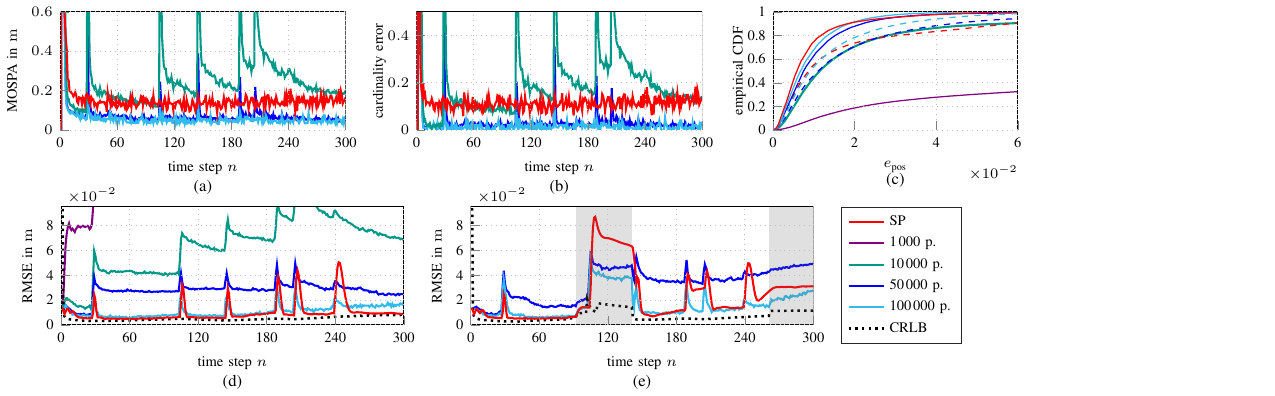}}
	\vspace*{-3mm}
	\caption{Simulation results in terms of the agent position \ac{rmse} from Ex.1 (d) and Ex.2 (e), as well as the mean OSPA of all VAs (a) and associated cardinality error (b) from Ex.1 over all time steps. (c) shows the eCDF of the agent position error for Ex.1 (full) and Ex.2 (dashed). Gray areas in (e) indicate \ac{olos} situations between agent and \ac{pa}.}\label{fig:results}
	\vspace{-3mm}
\end{figure*}

\subsection{Simulation Results}

\subsubsection*{Experiment 1}

In this experiment the agent and \ac{pa} have \ac{los} connection throughout the whole trajectory. The results are displayed in Fig.~\ref{fig:results}a - \ref{fig:results}d. The eCDF of the agent's position error in Fig.~\ref{fig:results}c and the \ac{rmse} in Fig.~\ref{fig:results}d show similar results for the proposed algorithm and particle-based \ac{mpslam} with $100\ist000$ particles, both of them almost reaching the CRLB. The mean OSPA of all VAs (Fig.~\ref{fig:results}a) is higher for the proposed algorithm when compared to the $50\ist000$ and $100\ist000$ particle versions, which can be attributed to a higher mean cardinality error displayed in Fig.~\ref{fig:results}b. The $10\ist000$-particle implementation leads to an agent estimation error larger than $10\mathrm{cm}$ in $8\%$ of cases. The differences in runtime are significant, with the proposed algorithm being about $100$ times faster than the particle-based \ac{mpslam} with $100\ist000$ particles and around $10$ times faster for $10\ist000$ particles. Comparable runtimes could be achieved using $1\ist000$ particles, which, however, leads to a total loss of the agent's trajectory in all $500$ realizations.
\renewcommand{\arraystretch}{1.2} 
\begin{table}[H]
	\caption{Mean runtime per iteration.}
	\label{tab:mimo_exec}   
	\begin{tabular}{|c|c|c|c|c|c|}
		\hline  
		SP		 & 1\ist000 p. & 10\ist000 p. & 50\ist000 p. & 100\ist000 p.\\ \hline
		$0.029\mathrm{s}$ & $0.039\mathrm{s}$ & $0.275\mathrm{s}$	& $0.948\mathrm{s}$    & $1.936\mathrm{s}$  \\ \hline
	\end{tabular}        
\end{table}
\subsubsection*{Experiment 2}
The scenario is displayed in Fig.~\ref{fig:floorplan}, where a wall obstructs the LOS connection to the PA as well as to some VAs over some parts of the trajectory. The best result is achieved by the particle-based \ac{mpslam} with $100\ist000$ particles, as displayed in Fig.~\ref{fig:results} (c) and (e). For the proposed algorithm the agent position is lost in $1\%$ of realizations, which were removed from the RMSE plot in Fig.~\ref{fig:results}(e). Execution times are in close correspondence to Ex. 1 and are listed in Tab.~\ref{tab:mimo_exec}.

\subsection{Measurement Results}
\begin{figure}[ht]
	\centering
	\scalebox{0.55}{\includegraphics[]{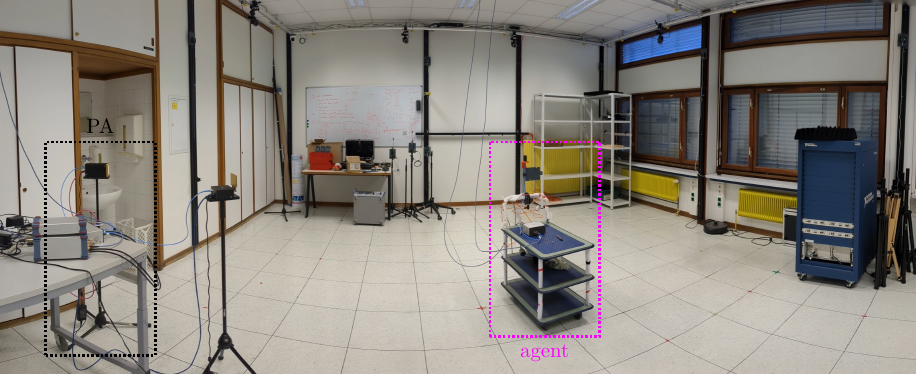}}
	\caption{Picture taken in the NXP laboratory room at TU Graz, showing the measurement setup with PA and agent.}
	\label{fig:nxplab}
\end{figure}

\begin{figure}[ht]
	\centering
	\hspace*{-2.4cm}
	\scalebox{1}{\includegraphics[]{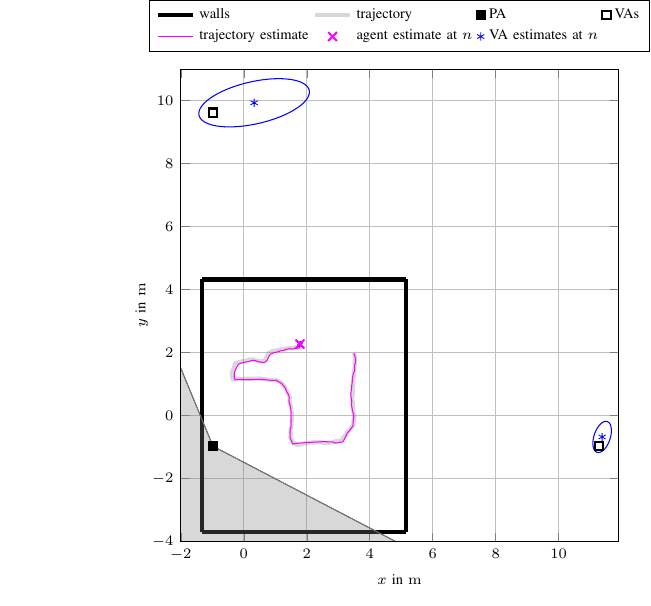}}
	\vspace*{-3mm}
	\caption{Floorplan of the NXP laboratory room used for measurements showing the room, agent trajectory and the \ac{pa} with its field of view and two \acp{va}. Overlaid, the estimated agent and VA positions are shown, along with a visualization of their covariance matrix (10-fold) at time $n=92$.}
	\label{fig:nxp_floorplan}
\end{figure}

Measurements were conducted in the NXP laboratory room at TU Graz shown in Fig.~\ref{fig:nxplab}, with one PA equipped with an antenna array and the agent having a single antenna, making it a multiple input, single output (MISO) scenario. Fig.~\ref{fig:nxp_floorplan} also provides the agent's trajectory, which consists of a total of $92$ steps spaced approximately $10~\mathrm{cm}$ apart. 
Reference measurements were taken using an optical motion capture system from Qualisys, which provides ground truth measurements with an accuracy in the order of millimetres. The \ac{pa} is equipped with a $4\times1$ phased-array with field of view of $\pm 45^{\circ}$ and a $3~\mathrm{dB}$ beamwidth of $25^{\circ}$, with the beam steered in steps of $2.5^{\circ}$. The limited field of view results in only two of the four walls being fully visible and parts of the room being invisible (see Fig.~\ref{fig:nxp_floorplan}). The agent is represented by an antenna with omnidirectional radiation pattern in the horizontal plane and negligible radiation in vertical direction, making ground and ceiling reflections unlikely. Measurements were made using an Ilmens M-sequence direct correlation channel sounder operating at a carrier frequency of $f_c=6.95\mathrm{GHz}$. The pulse shape is given by a raised-cosine pulse with a $3\mathrm{GHz}$ $3\mathrm{dB}$-bandwidth and a roll-off factor of $0.6$. A total of $318$ samples were used limiting the maximum observable distance to $20~\mathrm{m}$.
\begin{figure}[ht]
	\centering
	\hspace*{-2.2cm}
	\scalebox{1}{\includegraphics[]{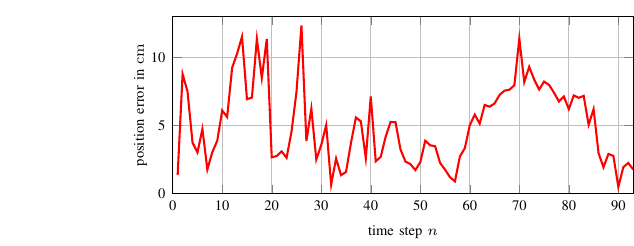}}
	\vspace{-1mm}
	\caption{Measurement results in terms of agent positioning error.}
	\label{fig:meas_res}
	\vspace{-6mm}
\end{figure}
\subsubsection*{Experiment 3}
The primary objective of this experiment was to verify the functionality of the algorithm using real-life measurement data, with an emphasis on qualitative rather than quantitative evaluation. 
The posterior map in Fig.~\ref{fig:nxp_floorplan} shows that both visible VAs were detected for parts of the trajectory. The agent position error is displayed in Fig.~\ref{fig:meas_res}, with the error being higher in the parts of the trajectory where the agent is near the edges of the field of view due to a decreased normalized amplitude.

\section{Conclusion}\label{sec:conclusion}

\acresetall
We proposed a low complexity implementation of the \ac{spa} algorithm for \ac{mpslam}. By using the uncented or \ac{sp} transform to approximate \acp{pdf} as Gaussian, integrals involved in the \ac{spa} can be efficiently evaluated and posterior \acp{pdf} accurately represented. This is particularly suitable for \ac{mimo} systems, where the joint availability of \ac{toa}, \ac{aoa} and \ac{aod} measurements leads to unambiguous transformations, allowing the resulting joint posterior \ac{pdf} to be approximated accurately by Gaussian densities.
Through numerical evaluation in two different \ac{mimo} settings, we demonstrated that the proposed algorithm achieves accurate and robust localization results with runtimes in the order of tens of milliseconds. In comparison, a particle-based \ac{mpslam} algorithm required a high number of particles to achieve similar localization performance, resulting in significantly increased runtimes.

\vspace{-1mm}
\appendix
\section{Message Approximation}\label{sec:app}
In this appendix, we discuss the \ac{sp}-based approximation of the SPA messages, with the derivation of their analytical counterparts displayed in the supplementary material of \cite{VenLeiTerMeyWit:TWC2024}. We adopt the same notation as in \cite{VenLeiTerMeyWit:TWC2024} and denote the approximated messages by adding a tilde symbol as $\tilde{A}$. At time $n-1$, the state of the agent $\V{x}_{n-1}$ and legacy PVAs $\V{y}_{n-1}^{(j)}=[\RV{\psi}^{(j)}_{k,n}{}^{\mathrm{T}} \;\rv{r}^{(j)}_{k,n}]^{\mathrm{T}}$ are assumed to follow Gaussian PDFs with mean vectors $\V{\hat{x}}_{n-1}$ and $\underline{\hat{\V{\psi}}}_{k,n-1}^{(j)}$ and covariance matrices $\V{P}_{n-1}$ and $\underline{\V{Q}}_{k,n-1}^{(j)}$ respectively. Here $\RV{\psi}^{(j)}_{k,n}$ is the \ac{pbo} position and $r_{n-1}$ the existence variable, with existence probability $p(r_{n-1}=1)\triangleq e_{k,n-1}^{(j)}$.

\subsubsection*{1. Prediction}
Applying the agent state-transition model from Sec.~\ref{sec:state_transition_model} yields 
\begin{align}
	&\tilde{\chi}_\text{x}( \V{x}_n) = f_\text{N}(\V{x}_n;\; \V{\hat{x}}_n^-, \V{P}_{n}^-)\hspace{4mm} \text{with}\label{eq:pred_agent}\nn\\
	&\V{\hat{x}}_n^- = \V{A}\V{\hat{x}}_{n-1}, \hspace{5mm} \V{P}_{n}^- = \V{A} \V{P}_{n-1} \V{A}^T + \bm{C}_{\text{x}}.
\end{align}
PVAs are affected by the survival probability $p_s$ as
\begin{align}
	\tilde{\chi}\big(  \underline{\V{\psi}}_{k,n}^{(j)}  \ist, 1\big) = p_{\mathrm{s}} e_{k,n-1}^{(j)} f_N(\underline{\V{\psi}}_{k,n}^{(j)};\; \underline{\hat{\V{\psi}}}_{k,n}^{(j)-}, \underline{\V{Q}}_{k,n}^{(j)-}) \label{eq:pred_va}\\[-7mm]\nn
\end{align}
with $\underline{\hat{\V{\psi}}}_{k,n}^{(j)-}\rmv=\rmv\underline{\hat{\V{\psi}}}_{k,n-1}^{(j)}$ and $\underline{\V{Q}}_{k,n}^{(j)-}\rmv=\rmv\underline{\V{Q}}_{k,n-1}^{(j)}$.
\subsubsection*{2a. Measurement Evaluation legacy PVAs} In the case $\underline{a}_{k,n}^{(j)}=0$ the message equals $\tilde{\beta}\big( \underline{a}_{k,n}^{(j)} \big) = (1-\pd) \chi_{k,n}^{(j)}$ with $\chi_{k,n}^{(j)} = (1-p_s e_{k,n-1}^{(j)})$ and otherwise \cite[Eq. 5]{VenLeiTerMeyWit:TWC2024}
\begin{align}
	&\beta\big( \underline{a}_{k,n}^{(j)} \big)\rmv\rmv\rmv = \frac{\pd}{\mu_\mathrm{fa} \fa} \rmv\rmv\rmv\int\!\!\! \rmv \rmv \int \rmv\rmv\chi_\text{x}( \V{x}_n) \chi\big(  \underline{\V{\psi}}_{k,n}^{(j)}  \ist, 1\big)\nn\\
	&\hspace{11mm}\times f\big(\V{z}_{m,n}^{(j)}| \V{x}_n, \underline{\V{\psi}}_{k,n}^{(j)}\big)\mathrm{d}\V{x}_{n}\mathrm{d} \underline{\V{\psi}}_{k,n}^{(j)}.\
\end{align}
This integral is solved using \acp{sp}, which results in the \ac{kf} innovation equation \cite[p. 202]{BarShalom2002EstimationTracking} and leads to
\begin{equation}\label{eq:beta_margi}
	\tilde{\beta}\big( \underline{a}_{k,n}^{(j)} \big) = \frac{p_{\mathrm{s}} \pd e_{k,n-1}^{(j)}}{\mu_\mathrm{fa} \fa} f_\text{N}(\V{z}_{m,n}^{(j)}; \; \underline{\V{\mu}}_{k, n}^{(j)},\; \V{C}_{\text{z},m} + \underline{\V{C}}_{k,n}^{(j)})
\end{equation}
where $\underline{\V{\mu}}_{k, n}^{(j)}$ and $\underline{\V{C}}_{k,n}^{(j)}$ are calculated as shown in \ref{sec:sp} and $\V{C}_{\text{z},m}$ is the covariance of the measurement noise. We denote the normal \ac{pdf} as a partial result $E_{k,n}^{(j)}=f_\text{N}(\V{z}_{m,n}^{(j)}; \; \underline{\V{\mu}}_{k, n}^{(j)},\; \V{C}_{\text{z},m} + \underline{\V{C}}_{k,n}^{(j)})$.
\subsubsection*{2b. Measurement Evaluation new PVAs}
For $\underline{a}_{k,n}^{(j)}\in\mathcal{K}^{(j)}_{n-1}$ the message equals $\xi\big(\overline{a}^{(j)}_{m,n}\big)=1$ \cite[Eq. 7]{VenLeiTerMeyWit:TWC2024} and for $\underline{a}_{k,n}^{(j)}=0$  
\begin{align}
	&\xi\big(\overline{a}^{(j)}_{m,n}\big)= 1 +\frac{ {\mu}_{\mathrm{n}} }{ \mu_\mathrm{fa} \ist \fa } 
	\int\!\!\!\int\rmv\rmv \chi_\text{x}( \V{x}_n)f_{\mathrm{n}}\big(\overline{\V{\psi}}_{m,n}^{(j)}\big) \nn \\
	&\hspace{13mm}\times  f\big(\V{z}_{m,n}^{(j)}| \V{x}_n, \overline{\V{\psi}}_{m,n}^{(j)}\big) \, \mathrm{d}\V{x}_{n} \ist \mathrm{d} \overline{\V{\psi}}_{m,n}^{(j)}.
\end{align}
New PVAs are assumed to follow a uniform PDF across all domains denoted as $f_{\mathrm{n}}\big(\overline{\V{\psi}}_{m,n}^{(j)}\big)\triangleq f_\text{U}(\overline{\V{\psi}}_{m,n}^{(j)})$. The outer integral is approximated as described in the appendix to \cite{KimGraSveKimWym:TVT2022} and entails performing importance sampling with $f_\text{U}(\overline{\V{\psi}}_{m,n}^{(j)})$ acting as target distribution. 

To compute the messages associated with the new PVA states $\overline{\V{\psi}}_{m,n}^{(j)}$ (i.e., equations \eqref{eq:evalnewVAsnoapprox}, \eqref{eq:existencenewpvas}, and \eqref{eq:newpvabelief}) accurately, direct sampling from $f_{\mathrm{U}}(\overline{\V{\psi}}_{m,n}^{(j)})$ requires too many samples and is computationally demanding. Hence, we instead draw samples from a suitable proposal density
\begin{align}
	f_{\mathrm{pr}}\big(\overline{\V{\psi}}_{m,n}^{(j)}\big) = f_\text{N}(\overline{\V{\psi}}_{m,n}^{(j)}; \overline{\hat{\V{\psi}}}_{m,n}^{(j)}, \overline{\V{Q}}_{m,n}^{(j)}) \label{eq:newVAKIM}\\[-6mm]\nn
\end{align}
which is calculated by transforming new measurements into the VA domain as follows. A set of SPs is selected for both agent state $\{(\tilde{\V{x}}_n^{(i)}, w_\text{m}^{(i)}, w_\text{c}^{(i)})\}_{i=0}^I$ and measurement state $\{(\tilde{\V{z}}_{m,n}^{(j,l)}, w_\text{m}^{(j,l)}, w_\text{c}^{(j,l)})\}_{l=0}^{L}$, with $I$ and $L$ denoting the number of SPs necessary to cover the respective state dimensionality. Then, each possible SP combination is transformed into the VA space via relations from Section \ref{sec:geometric_model}, yielding a set of $O=IL$ SPs associated with the distribution of new PVAs as $\tilde{\V{\psi}}_{m,n}^{(j,o)}=h(\V{\tilde{x}}_n^{(i)}, \tilde{\V{z}}_{m,n}^{(j,l)})$, where $h(\cdot)$ is the nonlinear function transforming into the VA domain. From the resulting SP set $\{(\overline{\V{\psi}}_{m,n}^{(j,o)}, w_\text{m}^{(o)}, w_\text{c}^{(o)})\}_{o=0}^{O=I\cdot L}$, the mean vector $\overline{\hat{\V{\psi}}}_{m,n}^{(j)}$ and covariance matrix $\overline{\V{Q}}_{m,n}^{(j)}$ are calculated as shown in \ref{sec:sp}.

The outer integral is approximated using importance sampling with $\mathrm{P}$ samples drawn from the proposal density $\overline{\V{\psi}}_{m,n,p}^{(j)}\sim f_{\mathrm{pr}}\big(\overline{\V{\psi}}_{m,n}^{(j)}\big)$, with corresponding weights $w_{m,n,p}^{(j)} \propto f_\text{U}(\overline{\V{\psi}}_{m,n}^{(j)})/ f_{\mathrm{pr}}\big(\overline{\V{\psi}}_{m,n,p}^{(j)}\big)$  leading to
\begin{align} 
	&\tilde{\xi}\big(\overline{a}^{(j)}_{m,n}\big) = 1 + \frac{ {\mu}_{\mathrm{n}} }{ \mu_\mathrm{fa} \ist \fa } \sum^{P}_{p=1} w_{m,n,p}^{(j)} \int\rmv\rmv  f_\text{N}(\V{x}_n;\; \V{\hat{x}}_n^-, \V{P}_{n}^-)\nn \\
	&\hspace*{15mm}\times f_{\mathrm{pr}}\big(\overline{\V{\psi}}_{m,n,p}^{(j)}\big) f\big(\V{z}_{m,n}^{(j)}| \V{x}_n, \overline{\V{\psi}}_{m,n,p}^{(j)}\big)\mathrm{d}\V{x}_n. \label{eq:evalnewVAsnoapprox}
\end{align}
In line with \eqref{eq:beta_margi} the inner integral is solved using \acp{sp}, which results in the \ac{kf} innovation equation \cite[p. 202]{BarShalom2002EstimationTracking}, leading to
\vspace*{-2mm}
\begin{align}
	\tilde{\xi}\big(\overline{a}^{(j)}_{m,n}\big) &= 1 + \frac{ {\mu}_{\mathrm{n}} }{ \mu_\mathrm{fa} \ist \fa } \sum^{P}_{p=1} w_{m,n,p}^{(j)} \nn\\
	&\hspace*{5mm}\times f_\text{N}(\V{z}_{m,n}^{(j)}; \; \overline{\V{\mu}}_{m, n, p}^{(j)},\; \V{C}_{\text{z},m} + \overline{\V{C}}_{m,n,p}^{(j)} )\label{eq:evalnewVAs}
\end{align}
where $\overline{\V{\mu}}_{m, n, p}^{(j)}$ and $\overline{\V{C}}_{m,n,p}^{(j)}$ are calculated as shown in \ref{sec:sp}. 

\subsubsection*{3. Loopy Data Association} Messages $\tilde{\beta}\big( \underline{a}_{k,n}^{(j)} \big)$ and $\tilde{\xi}\big(\overline{a}^{(j)}_{m,n}\big)$ are used for the loopy DA to calculate the approximate messages $\tilde{\eta}\big( \underline{a}_{k,n}^{(j)} \big)$ and $\tilde{\varsigma}\big( \overline{a}_{m,n}^{(j)}\big)$ according to \cite{WilLau:J14}.
\subsubsection*{4a. Existence of legacy PVAs}
The existence of legacy PVAs is determined as 
\begin{align}\label{eq:existance_legacy}
	\vspace*{-1mm}
	e_{k,n}^{(j)} &= p_{\mathrm{s}} e_{k,n-1}^{(j)} \tilde{\eta}\big(\underline{a}_{k,n}^{(j)}\!\rmv=\!0\big) (1 \!-\rmv \pd \ist) + \frac{p_{\mathrm{s}} e_{k,n-1}^{(j)} \pd \ist  }  {\mu_\mathrm{fa} \ist \fa}\nn\\
	&\times\rmv\rmv \sum^{M_n^{(j)}}_{\underline{a}_{k,n}^{(j)} = 1} \!\! \tilde{\eta}\big( \underline{a}_{k,n}^{(j)} \big) f_\text{N}(\V{z}_{m,n}^{(j)}; \; \underline{\V{\mu}}_{k, n}^{(j)},\; \V{C}_{\text{z}} + \underline{\V{C}}_{k,n}^{(j)}).
\end{align}
Note that in \eqref{eq:existance_legacy}, the Gaussian \ac{pdf} $f_\text{N}(\cdot)$ corresponds to $E_{k,n}^{(j)}$. 
\subsubsection*{4b. Existence of new \acp{pbo}}
 The existence of new PVAs is determined as 
\begin{align}
	\vspace*{-1mm}
	e_{m,n}^{(j)} &= \tilde{\varsigma}\big( \overline{a}_{m,n}^{(j)} \!\rmv=\! 0\big) \frac{\mu_{\mathrm{n}} }{\mu_\mathrm{fa} \ist \fa}\sum^{P}_{p=1} w_{m,n,p}^{(j)} \nn\\
	&\hspace*{1mm}\times f_\text{N}(\V{z}_{m,n}^{(j)}; \; \overline{\V{\mu}}_{m, n, p}^{(j)},\; \V{C}_{\text{z},m} + \overline{\V{C}}_{m,n,p}^{(j)} ) + \phi^{(j)}_{m,n} \, . \label{eq:existencenewpvas}
\end{align}
Note that in \eqref{eq:existencenewpvas} the Gaussian  \ac{pdf} $f_\text{N}(\cdot)$ corresponds to \eqref{eq:evalnewVAs} and $\phi_{m,n}^{(j)} \triangleq\ist \tilde{\phi}\big( \overline{\V{\psi}}^{(j)}_{m,n},0 \big) =\rmv\sum^{K_{n-1}^{(j)}}_{\overline{a}_{m,n}^{(j)} = 0} \!\! \tilde{\varsigma}\big( \overline{a}_{m,n}^{(j)}\big)$. 
\subsubsection*{5a. Agent Belief}
The agent belief \cite[Eq. 18]{VenLeiTerMeyWit:TWC2024} is calculated by inserting \cite[Eq. 13]{VenLeiTerMeyWit:TWC2024}, which leads to
\begin{align}
	q(\V{x}_n)\rmv &=\frac{1}{C_{\text{x}n}}\rmv\prod_{j=1}^J \rmv\rmv\rmv \prod_{k \in \Set{K}_{n-1}^{(j)}}\rmv\rmv\rmv\rmv\rmv\rmv A_{k,n}^{(j)} \chi_\text{x}( \V{x}_n) + B_{k,n}^{(j)}\rmv\rmv\rmv\rmv\sum^{M_n^{(j)}}_{\underline{a}_{k,n}^{(j)} = 1} \!\! \tilde{\eta}\big( \underline{a}_{k,n}^{(j)} \big) \nn\\
	&\hspace*{1mm}\times \rmv\rmv\int\rmv\rmv  \chi_\text{x}( \V{x}_n) \chi\big(  \underline{\V{\psi}}_{k,n}^{(j)}  \ist, 1\big)f\big(\V{z}_{m,n}^{(j)}| \V{x}_n, \underline{\V{\psi}}_{k,n}^{(j)}\big) \mathrm{d} \underline{\V{\psi}}_{k,n}^{(j)}
\end{align}
where the normalization constant factor $C_{\text{x}n}$ can be disregarded, as the final distribution has to follow a Gaussian \ac{pdf} and the terms $A_{k,n}^{(j)}=\tilde{\eta}\big(\underline{a}_{k,n}^{(j)} \!\rmv=\! 0\big) \big[\chi_{k,n}^{(j)} + (1 \!-\rmv \pd \ist) p_{\mathrm{s}} e_{k,n-1}^{(j)}]$ and $B_{k,n}^{(j)}= (\pd \ist p_{\mathrm{s}} e_{k,n-1}^{(j)})/ (\mu_\mathrm{fa} \ist \fa)$ are introduced for brevity. The integral is computed considering the joint Gaussian distribution defined by the mean vector $\V{\mu}_{k,n}^{-(j)}=[\V{\hat{x}}_n^{-\text{T}}\underline{\hat{\V{\psi}}}_{k,n}^{(j)-\ist\text{T}}]^{\text{T}}$ and covariance matrix $\V{C}_{k,n}^{(j)}=\bdiag{\{\V{P}_{n}^-,\underline{\V{Q}}_{k,n}^{(j)-}\}}$ using \acp{sp}, which results in a \ac{kf} update for both the agent and \ac{pbo} $k$ of anchor $j$ as
\vspace*{-2mm}
\begin{align}
	\V{\mu}_{k,m,n}^{(j)} &= \V{\mu}_{k,n}^{-(j)} + \V{K}_{m,n}^{(j)} (\V{z}_{m,n}-\tilde{\V{\mu}}_{k,n}^{-(j)})\\
	\V{C}_{k,m,n}^{(j)} &= \V{C}_{k,n}^{-(j)}-\V{K}_{m,n}^{(j)}(\tilde{\V{C}}_{k,n}^{-(j)} + \V{C}_{\text{z},m})\V{K}_{m,n}^{(j)\text{T}}\\[-7mm]\nn
\end{align}
where $\V{K}_{m,n}^{(j)} = \tilde{\V{C}}_{\text{z},k,n}^{-(j)} (\tilde{\V{C}}_{k,n}^{-(j)}+ \V{C}_{\text{z},m})^{-1}$ is the Kalman gain, and $\tilde{\V{\mu}}_{k,n}^{-(j)}$, $\tilde{\V{C}}_{k,n}^{-(j)}$ and $\tilde{\V{C}}_{\text{z},m,k,n}^{-(j)}$ result from the \ac{sp}-transform. The mean $\V{\hat{x}'}_{m,n}$ and covariance matrix $\V{P}_{m,n}'$ are recovered from $\V{\mu}_{k,m,n}^{(j)}$ and $\V{C}_{k,m,n}^{(j)}$ (ignoring the block-crossvariance matrices) leading to
\begin{align}\label{eq:bel1}
	\tilde{q}(\V{x}_n) &=\prod_{j=1}^J \prod_{k \in \Set{K}_{n-1}^{(j)}} A_{k,n}^{(j)} f_\text{N}(\V{x}_n;\; \V{\hat{x}}_n^-, \V{P}_{n}^-) + B_{k,n}^{(j)}\nonumber\\
	&\hspace*{-2mm}\times \sum^{M_n^{(j)}}_{\underline{a}_{k,n}^{(j)} = 1} \!\! \tilde{\eta}\big( \underline{a}_{k,n}^{(j)} \big) E_{k,n}^{(j)}f_\text{N}(\V{x}_n;\; \V{\hat{x}'}_{m,n}, \V{P}_{m,n}').\\[-7mm]\nn
\end{align}
Since the Kalman update provides the posterior \ac{pdf}, the evidence term needs to be accounted for as stated in Bayes' theorem, i.e. the resulting distribution has to be multiplied with $E_{k,n}^{(j)}$ from \eqref{eq:beta_margi}. Since the weighted sum of Gaussian \acp{pdf} in \eqref{eq:bel1} is not a Gaussian distribution itself, it is approximated using moment matching \cite[p. 55]{BarShalom2002EstimationTracking}, yielding a Gaussian distribution with mean $\V{\hat{x}}_{k,n}^{(j)}$ and covariance matrix $\V{P}_{k,n}^{(j)}$. Finally, neglecting the normalization constant, the product of Gaussian \acp{pdf} is determined by \cite{bromiley2003products}
\vspace*{-1mm}
\begin{align}
	\tilde{q}(\V{x}_n) &=\prod_{j=1}^J \prod_{k \in \Set{K}_{n-1}^{(j)}} f_\text{N}(\V{x}_n;\; \V{\hat{x}}_{k,n}^{(j)}, \V{P}_{k,n}^{(j)})\nonumber\\
	&\propto f_\text{N}(\V{x}_n;\; \V{\hat{x}}_n, \V{P}_n)\\[-7mm]\nn
\end{align}
where $\V{P}_n = \big(\sum_{j=1}^{J} \sum_{k \in \Set{K}_{n-1}^{(j)}} \V{P}_{k,n}^{-1(j)}\big)^{-1}$ and $\V{\hat{x}}_n= \V{P}_n\sum_{j=1}^{J} \sum_{k \in \Set{K}_{n-1}^{(j)}} \V{P}_{k,n}^{-1(j)}\V{\hat{x}}_{k,n}^{(j)}$.
\subsubsection*{5b. Legacy PVAs belief}
The \ac{pbo} belief \cite[Eq. 19]{VenLeiTerMeyWit:TWC2024} is calculated as
\begin{align}
	{q}\big( \underline{\V{\psi}}_{k,n}^{(j)}, 1\big) = \frac{1}{\underline{C}_{k,n}^{(j)}} \chi\big( \underline{\V{\psi}}_{k,n}^{(j)} , 1\big) \gamma\big(  \underline{\V{\psi}}_{k,n}^{(j)},  1 \big)
\end{align}
and approximated neglecting the normalization constant $C_{k,n}^{(j)}$. Plugging in Eq.~\eqref{eq:pred_va} and the measurement update message $\gamma\big(  \underline{\V{\psi}}_{k,n}^{(j)},  1 \big)$ \cite[Eq. 14]{VenLeiTerMeyWit:TWC2024}, leads to
\begin{align}
	\tilde{q}\big( \underline{\V{\psi}}_{k,n}^{(j)}, 1\big) &= p_{\mathrm{s}} e_{k,n-1}^{(j)} f_\text{N}(\underline{\V{\psi}}_{k,n}^{(j)};\; \underline{\hat{\V{\psi}}}_{k,n}^{(j)-}, \underline{\V{Q}}_{k,n}^{(j)-}) \tilde{\eta}\big(\underline{a}_{k,n}^{(j)}\!\rmv=\!0\big)\nn\\
	&\times (1 \!-\rmv \pd \ist)+\frac{ \pd p_{\mathrm{s}} e_{k,n-1}^{(j)} }  {\mu_\mathrm{fa} \ist \fa} \sum^{M_n^{(j)}}_{\underline{a}_{k,n}^{(j)} = 1} \!\! \tilde{\eta}\big( \underline{a}_{k,n}^{(j)} \big)\nn\\
	&\times E_{k,n}^{(j)}f_\text{N}(\underline{\V{\psi}}_{k,n}^{(j)};\; \underline{\hat{\V{\psi}}'}_{k,n}^{(j)}, \underline{\V{Q}'}_{k,n}^{(j)})\nn\\
	&\approx f_\text{N}(\underline{\V{\psi}}_{k,n}^{(j)};\; \underline{\hat{\V{\psi}}}_{k,n}^{(j)}, \underline{\V{Q}}_{k,n}^{(j)})
\end{align}
where $\underline{\hat{\V{\psi}}'}_{k,n}^{(j)}$ and $\underline{\V{Q}'}_{k,n}^{(j)}$ result from the KF update, the partial result $E_{k,n}^{(j)}$ is given in \eqref{eq:beta_margi} and the sum is approximated again using using moment matching \cite[p. 55]{BarShalom2002EstimationTracking}.
\subsubsection*{5c. New PVAs belief} For new PVAs \cite[Eq. 21]{VenLeiTerMeyWit:TWC2024}
\begin{align}
	{q}\big( \overline{\V{\psi}}_{m,n}^{(j)}, 1\big) = \frac{1}{\overline{C}_{m,n}^{(j)}} \phi\big( \overline{\V{\psi}}^{(j)}_{m,n}, 1 \big) \label{eq:newpvabelief}
\end{align}
 the proposal density from Eq.~\eqref{eq:newVAKIM} is used as distribution for all new PVAs as
\begin{align}
	\tilde{q}\big( \overline{\V{\psi}}_{m,n}^{(j)}, 1\big)&=f_{\text{N}}(\overline{\V{\psi}}_{m,n}^{(j)}; \overline{\hat{\V{\psi}}}_{m,n}^{(j)}, \overline{\V{Q}}_{k,n}^{(j)}) 
\end{align}
in accordance with \eqref{eq:existencenewpvas}.


%

 
 \acrodef{mimo}[MIMO]{multiple input multiple output}
 \acrodef{awgn}[AWGN]{additive white Gaussian noise}
 \acrodef{bw}[BW]{bandwidth}
 \acrodef{blt}[BLT]{bluetooth}
 \acrodef{cdf}[CDF]{cumulative distribution function}
 \acrodef{crlb}[CRLB]{Cram\'er-Rao lower bound}
 \acrodef{dmc}[DMC]{dense multipath component}
 \acrodef{dut}[DUT]{device under test}
 \acrodef{eirp}[EIRP]{equivalent isotropic radiated power}
 \acrodefplural{esl}[ESLs]{electronic shelf labels} 
 \acrodef{los}[LOS]{line-of-sight}
 \acrodef{mf}[MF]{matched filter}
 \acrodef{ml}[ML]{maximum likelihood}
 \acrodef{mpc}[MPC]{multipath component}
 \acrodef{nlos}[NLOS]{non-line-of-sight}
 \acrodef{pcb}[PCB]{printed circuit board}
 \acrodef{pdf}[PDF]{probability density function}
 \acrodef{reb}[REB]{ranging error bound}
 \acrodef{rss}[RSS]{received signal strength}
 \acrodef{smc}[SMC]{specular multipath component}
 \acrodef{snr}[SNR]{signal-to-noise-ratio}
 \acrodef{sinr}[SINR]{signal-to-interference-plus-noise-ratio}
 \acrodef{tdoa}[TDOA]{time difference of arrival}
 \acrodef{tka}[TKA]{trusted keyless access}
 \acrodef{toa}[TOA]{time-of-arrival}
 \acrodef{aoa}[AOA]{angle-of-arrival}
 \acrodef{aod}[AOD]{angle-of-departure}
 \acrodef{uwb}[UWB]{ultra wide band}
 \acrodef{mie}[MIE]{method of interval estimation}
 \acrodef{mc}[MC]{Monte Carlo}
 \acrodef{mse}[MSE]{mean squared error}
 \acrodef{ci}[CI]{confidence interval}
 \acrodef{cl}[CL]{confidence level}
 \acrodef{pdp}[PDP]{power delay profile}
 \acrodef{dps}[DPS]{delay power spectrum}
 \acrodef{dm}[DM]{dense multipath}
 \acrodef{nlike}[NLIKE]{normalized likelihood}
 \acrodef{zzb}[ZZB]{Ziv-Zakai bound}
 \acrodef{ut}[UT]{unscented transform}
 \acrodef{glrt}[GLRT]{generalized likelihood ratio test}
 \acrodef{mse}[MSE]{mean squared error}
 \acrodef{rmse}[RMSE]{root mean squared error}
 \acrodef{nnlike}[NNLIKE]{normalized noise-free likelihood}
 \acrodef{stdv}[STDV]{standard deviation}
 \acrodef{rv}[RV]{random variable}
 \acrodef{bp}[BP]{belief propagation}
 \acrodef{pda}[PDA]{probabilistic data association}
 \acrodef{mp}[MP]{multipath}
 \acrodef{pmf}[PMF]{probability mass function}
 \acrodef{pdaf}[PDAF]{probabilistic data association filter}
 \acrodef{pdaai}[AIPDA]{amplitude-information \ac{pda}}
 \acrodef{olos}[OLOS]{obstructed line-of-sight}
 \acrodef{spa}[SPA]{sum-product algorithm}
 \acrodef{fg}[FG]{factor graph}
 \acrodef{mmse}[MMSE]{minimum mean-square error}
 \acrodef{lhf}[LHF]{likelihood function}
 \acrodef{fa}[FA]{false alarm}
 \acrodef{ceda}[CEDA]{channel estimation and detection algorithm} 
 \acrodef{pcrlb}[P-CRLB]{posterior Cram\'er-Rao lower bound}
 \acrodef{mpslam}[MP-SLAM]{multipath-based simultaneous localization and mapping}
 \acrodef{va}[VA]{virtual anchor}
 \acrodef{dnr}[DNR]{dense-to-noise ratio}
 \acrodef{pbo}[PVA]{potential virtual anchor}
 \acrodef{npbo}[NPBO]{new \ac{pbo}}
 \acrodef{lpbo}[LPBO]{legacy \ac{pbo}}
 \acrodef{aednn}[AE-DNN]{autoencoder deep neural network}   
 \acrodef{gpr}[GPR]{Gaussian process regression}  
 \acrodef{cluster}[CLUSTER]{{\color{red}error}}  
 \acrodef{delaybias}[ML-BIAS]{{\color{red}error}}  
 \acrodef{gptrack}[GP-TRACK]{{\color{red}error}}  
 \acrodef{chslam}[CH-SLAM]{{\color{red}error}}  
 \acrodef{wrt}[w.r.t.]{with respect to} 
 \acrodef{pa}[PA]{physical anchor} 
  \acrodef{bs}[BS]{base station} 
 \acrodef{kf}[KF]{Kalman Filter}
 \acrodef{sp}[SP]{sigma point}



\vspace{-2mm} 
\bibliographystyle{IEEEtran}
\bibliography{IEEEabrv,references}

\begin{thebibliography}{10}
\providecommand{\url}[1]{#1}
\csname url@samestyle\endcsname
\providecommand{\newblock}{\relax}
\providecommand{\bibinfo}[2]{#2}
\providecommand{\BIBentrySTDinterwordspacing}{\spaceskip=0pt\relax}
\providecommand{\BIBentryALTinterwordstretchfactor}{4}
\providecommand{\BIBentryALTinterwordspacing}{\spaceskip=\fontdimen2\font plus
\BIBentryALTinterwordstretchfactor\fontdimen3\font minus
  \fontdimen4\font\relax}
\providecommand{\BIBforeignlanguage}[2]{{%
\expandafter\ifx\csname l@#1\endcsname\relax
\typeout{** WARNING: IEEEtran.bst: No hyphenation pattern has been}%
\typeout{** loaded for the language `#1'. Using the pattern for}%
\typeout{** the default language instead.}%
\else
\language=\csname l@#1\endcsname
\fi
#2}}
\providecommand{\BIBdecl}{\relax}
\BIBdecl

\bibitem{WitMeiLeiSheGusTufHanDarMolConWin:J16}
K.~Witrisal, P.~Meissner, E.~Leitinger, Y.~Shen, C.~Gustafson, F.~Tufvesson,
  K.~Haneda, D.~Dardari, A.~F. Molisch, A.~Conti, and M.~Z. Win,
  ``High-accuracy localization for assisted living: {5G} systems will turn
  multipath channels from foe to friend,'' \emph{{IEEE} Signal Process. Mag.},
  vol.~33, no.~2, pp. 59--70, Mar. 2016.

\bibitem{GentnerTWC2016}
C.~Gentner, T.~Jost, W.~Wang, S.~Zhang, A.~Dammann, and U.~C. Fiebig,
  ``Multipath assisted positioning with simultaneous localization and
  mapping,'' \emph{{IEEE} Trans. Wireless Commun.}, vol.~15, no.~9, pp.
  6104--6117, Sept. 2016.

\bibitem{KimGraSveKimWym:TVT2022}
H.~Kim, K.~Granstr{\"o}m, L.~Svensson, S.~Kim, and H.~Wymeersch, ``{PMBM-based
  SLAM} filters in {5G} {mmWave} vehicular networks,'' \emph{{IEEE} Trans. Veh.
  Technol.}, pp. 1--1, May 2022.

\bibitem{LeiMeyHlaWitTufWin:J19}
E.~{Leitinger}, F.~{Meyer}, F.~{Hlawatsch}, K.~{Witrisal}, F.~{Tufvesson}, and
  M.~Z. {Win}, ``A belief propagation algorithm for multipath-based {SLAM},''
  \emph{{IEEE} Trans. Wireless Commun.}, vol.~18, no.~12, pp. 5613--5629, Dec.
  2019.

\bibitem{LeiVenTeaMey:TSP2023}
E.~{Leitinger}, A.~{Venus}, B.~{Teague}, and F.~{Meyer}, ``Data fusion for
  multipath-based {SLAM}: {Combining} information from multiple propagation
  paths,'' \emph{{IEEE} Trans. Signal Process.}, vol.~71, pp. 4011--4028, Sep.
  2023.

\bibitem{MonThrKolWeg:AAAI2002}
M.~Montemerlo, S.~Thrun, D.~Koller, and B.~Wegbreit, ``{FastSLAM: A factored
  solution to the simultaneous localization and mapping problem},'' in
  \emph{Proc. AAAI-02}, Edmonton, Canda, Jul. 2002, pp. 593--598.

\bibitem{DurrantWhyte2006}
H.~Durrant-Whyte and T.~Bailey, ``Simultaneous localization and mapping: {Part
  I},'' \emph{IEEE Robot. Autom. Mag.}, vol.~13, no.~2, pp. 99--110, Jun. 2006.

\bibitem{LeiGreWit:ICC2019}
E.~{Leitinger}, S.~{Grebien}, and K.~{Witrisal}, ``Multipath-based {SLAM}
  exploiting {AoA} and amplitude information,'' in \emph{Proc. IEEE ICCW-19},
  Shanghai, China, May 2019, pp. 1--7.

\bibitem{MenMeyBauWin:J19}
R.~{Mendrzik}, F.~{Meyer}, G.~{Bauch}, and M.~Z. {Win}, ``Enabling situational
  awareness in millimeter wave massive {MIMO} systems,'' \emph{{IEEE} J. Sel.
  Topics Signal Process.}, vol.~13, no.~5, pp. 1196--1211, Sep. 2019.

\bibitem{KimGraGaoBatKimWym:TWC2020}
H.~{Kim}, K.~{Granstr{\"o}m}, L.~{Gao}, G.~{Battistelli}, S.~{Kim}, and
  H.~{Wymeersch}, ``{5G} {mmWave} cooperative positioning and mapping using
  multi-model {PHD} filter and map fusion,'' \emph{{IEEE} Trans. Wireless
  Commun.}, vol.~19, no.~6, pp. 3782--3795, Mar. 2020.

\bibitem{LiLeiCaiTuf:ICC2024}
X.~Li, X.~Cai, E.~Leitinger, and F.~Tufvesson, ``A belief propagation algorithm
  for multipath-based {SLAM} with multiple map features: {A} mmwave {MIMO}
  application,'' in \emph{Proc. IEEE ICC 2024}, Aug. 2024, pp. 269--275.

\bibitem{WieVenWilWitLei:Fusion2024}
L.~Wielandner, A.~Venus, T.~Wilding, K.~Witrisal, and E.~Leitinger, ``{MIMO}
  multipath-based {SLAM} for non-ideal reflective surfaces,'' in \emph{Proc.
  Fusion-2024}, Venice, Italy, Jul. 2024.

\bibitem{AruMasGorCla:TSP2002}
M.~S. Arulampalam, S.~Maskell, N.~Gordon, and T.~Clapp, ``A tutorial on
  particle filters for online nonlinear/non-{Gaussian} {Bayesian} tracking,''
  \emph{{IEEE} Trans. Signal Process.}, vol.~50, no.~2, pp. 174--188, Feb.
  2002.

\bibitem{Julier2004}
S.~J. Julier and J.~K. Uhlmann, ``Unscented filtering and nonlinear
  estimation,'' \emph{Proc. {IEEE}}, vol.~92, no.~3, pp. 401--422, Mar. 2004.

\bibitem{ArasaratnamHaykin2009_Cubature}
I.~Arasaratnam and S.~Haykin, ``Cubature {K}alman filters,'' \emph{{IEEE}
  Trans. Autom. Control}, vol.~54, no.~6, pp. 1254--1269, 2009.

\bibitem{Meyer2014_SigmaPointBP}
F.~Meyer, O.~Hlinka, and F.~Hlawatsch, ``Sigma point belief propagation,''
  \emph{{IEEE} Signal Process. Lett.}, vol.~21, no.~2, pp. 145--149, 2014.

\bibitem{Hansen2014}
T.~L. Hansen, M.~A. Badiu, B.~H. Fleury, and B.~D. Rao, ``A sparse {B}ayesian
  learning algorithm with dictionary parameter estimation,'' in \emph{Proc.
  IEEE SAM-14}, 2014, pp. 385--388.

\bibitem{HanFleRao:TSP2018}
T.~L. Hansen, B.~H. Fleury, and B.~D. Rao, ``Superfast line spectral
  estimation,'' \emph{{IEEE} Trans. Signal Process.}, vol.~PP, no.~99, pp. 2511
  -- 2526, Feb. 2018.

\bibitem{GreLeiWitFle:TWC2024}
S.~Grebien, E.~Leitinger, K.~Witrisal, and B.~H. Fleury, ``Super-resolution
  estimation of {UWB} channels including the dense component -- {An
  SBL}-inspired approach,'' \emph{{IEEE} Trans. Wireless Commun.}, vol.~23,
  no.~8, pp. 10\,301--10\,318, Feb. 2024.

\bibitem{Moederl2025}
J.~M\"oderl, A.~M. Westerkam, and E.~Leitinger, ``A block-sparse {B}ayesian
  learning algorithm with dictionary parameter estimation for multi-sensor data
  fusion,'' submitted to Fusion 2025, Jul. 7--11, 2025, Rio de Janeiro, Brazil.

\bibitem{LiLeiVenTuf:TWC2022}
X.~Li, E.~Leitinger, A.~Venus, and F.~Tufvesson, ``Sequential detection and
  estimation of multipath channel parameters using belief propagation,''
  \emph{{IEEE} Trans. Wireless Commun.}, pp. 1--1, Apr. 2022.

\bibitem{VenLeiTerMeyWit:TWC2024}
A.~{Venus}, E.~{Leitinger}, S.~{Tertinek}, F.~{Meyer}, and K.~{Witrisal},
  ``Graph-based simultaneous localization and bias tracking,'' \emph{{IEEE}
  Trans. Wireless Commun.}, vol.~23, no.~10, pp. 13\,141--13\,158, May 2024.

\bibitem{MeyKroWilLauHlaBraWin:J18}
F.~Meyer, T.~Kropfreiter, J.~L. Williams, R.~Lau, F.~Hlawatsch, P.~Braca, and
  M.~Z. Win, ``Message passing algorithms for scalable multitarget tracking,''
  \emph{Proc. {IEEE}}, vol. 106, no.~2, pp. 221--259, Feb. 2018.

\bibitem{LeiWieVenAsilomar2024_CoopSLAM}
E.~Leitinger, L.~Wielandner, A.~Venus, and K.~Witrisal, ``Multipath-based
  {SLAM} with cooperation and map fusion in {MIMO} systems,'' in \emph{Proc.
  Asilomar-24}, Pacifc Grove, CA, USA, Oct. 2024.

\bibitem{WilLau:J14}
J.~Williams and R.~Lau, ``Approximate evaluation of marginal association
  probabilities with belief propagation,'' \emph{IEEE Trans. Aerosp. Electron.
  Syst.}, vol.~50, no.~4, pp. 2942--2959, Oct. 2014.

\bibitem{Kay1993}
S.~M. Kay, \emph{Fundamentals of Statistical Signal Processing: Estimation
  Theory}.\hskip 1em plus 0.5em minus 0.4em\relax Upper Saddle River, NJ, USA:
  Prentice Hall, 1993.

\bibitem{KscFreLoe:TIT2001}
F.~Kschischang, B.~Frey, and H.-A. Loeliger, ``Factor graphs and the
  sum-product algorithm,'' \emph{{IEEE} Trans. Inf. Theory}, vol.~47, no.~2,
  pp. 498--519, Feb. 2001.

\bibitem{Loe:SMP2004_FG}
H.-A. Loeliger, ``An introduction to factor graphs,'' \emph{{IEEE} Signal
  Process. Mag.}, vol.~21, no.~1, pp. 28--41, Feb. 2004.

\bibitem{SchVoVo:TSP2008}
D.~Schuhmacher, B.-T. Vo, and B.-N. Vo, ``{A consistent metric for performance
  evaluation of multi-object filters},'' \emph{{IEEE} Trans. Signal Process.},
  vol.~56, no.~8, pp. 3447--3457, Aug. 2008.

\bibitem{TichavskyTSP1998}
P.~Tichavsky, C.~Muravchik, and A.~Nehorai, ``Posterior {Cramer-Rao} bounds for
  discrete-time nonlinear filtering,'' \emph{{IEEE} Trans. Signal Process.},
  vol.~46, no.~5, pp. 1386--1396, May 1998.

\bibitem{LeitingerJSAC2015}
E.~Leitinger, P.~Meissner, C.~Rudisser, G.~Dumphart, and K.~Witrisal,
  ``Evaluation of position-related information in multipath components for
  indoor positioning,'' \emph{{IEEE} J. Sel. Areas Commun.}, vol.~33, no.~11,
  pp. 2313--2328, Nov. 2015.

\bibitem{KalGeTalWymVal:Fusion2021}
O.~Kaltiokallio, Y.~Ge, J.~Talvitie, H.~Wymeersch, and M.~Valkama, ``{mmWave}
  simultaneous localization and mapping using a computationally efficient
  {EK-PHD} filter,'' in \emph{Proc. IEEE Fusion 2021}, Nov. 2021, pp. 1--8.

\bibitem{BarShalom2002EstimationTracking}
Y.~Bar-Shalom, T.~Kirubarajan, and X.-R. Li, \emph{Estimation with Applications
  to Tracking and Navigation}.\hskip 1em plus 0.5em minus 0.4em\relax New York,
  NY, USA: John Wiley \& Sons, Inc., 2002.

\bibitem{bromiley2003products}
\BIBentryALTinterwordspacing
P.~Bromiley, ``Products and convolutions of {G}aussian probability density
  functions,'' 2003. [Online]. Available: \url{leimao.github.io}
\BIBentrySTDinterwordspacing

\end{thebibliography}


\clearpage





	
\end{document}